\documentclass[twocolumn]{IEEEtran}
\usepackage{cite}
\usepackage[cmex10]{amsmath}
\usepackage{array}
\usepackage{amsfonts}
\usepackage{mathrsfs}
\usepackage{arydshln}
\usepackage{slashbox}
\usepackage{graphicx}
\usepackage{float}

\usepackage{subfigure}
\usepackage{amssymb}
\usepackage{amsmath}
\usepackage{enumitem}
\usepackage[colorlinks,linkcolor=black,urlcolor=black,anchorcolor=black,citecolor=black,hyperfootnotes=true]{hyperref}
\usepackage{multirow}
\usepackage{multicol}
\usepackage{cite}
\usepackage[subfigure]{graphfig}
\usepackage{xcolor}
\usepackage{setspace}
\newcommand{\mv}[1]{\mbox{\boldmath{$ #1 $}}}
\usepackage[linesnumbered,ruled]{algorithm2e}
\newtheorem{theorem}{\underline{Theorem}}
\newtheorem{assumption}{\underline{Assumption}}
\newtheorem{lemma}{\underline{Lemma}}
\newtheorem{proposition}{\underline{Proposition}}

\newcommand{\qed}{\nobreak \ifvmode \relax \else
	\ifdim\lastskip<1.5em \hskip-\lastskip
	\hskip1.5em plus0em minus0.5em \fi \nobreak
	\vrule height0.75em width0.5em depth0.25em\fi}
{\setlength\abovedisplayskip{0.4pt}
	\setlength\belowdisplayskip{0.4pt}
	\newcommand{\cc}{\mathrm{C}}
	\newcommand{\dd}{\mathrm{D}}
	\newcommand{\OO}{\mathrm{O}}
	\newcommand{\II}{\mathrm{I}}
	\newcommand{\ttt}{\mathrm{T}}
	\newcommand{\ff}{\mathrm{F}}
	\newcommand{\ub}{\mathrm{U}}
	\newcommand{\BC}{\mathrm{BC}}
	\newcommand{\PP}{\mathrm{P}}
	\setlength{\textfloatsep}{0.36pt}
	\setlength{\skip\footins}{7pt}
	
\begin{document}
		\bstctlcite{bstctl:etal, bstctl:nodash, bstctl:simpurl}
		\title{{Intelligent Reflecting Surface Aided\\ Multi-User Communication: \\Capacity Region and Deployment Strategy}
			\author{Shuowen Zhang and Rui Zhang}\thanks{This paper was presented in part at the IEEE International Workshop on Signal Processing Advances for \hbox{Wireless Communications (SPAWC), 2020 \cite{SPAWC}.}}\thanks{S. Zhang is with the Department of Electronic and Information Engineering, The Hong Kong Polytechnic University (e-mail: shuowen.zhang@polyu.edu.hk). She was with the Department of Electrical and Computer Engineering, National University of Singapore. R. Zhang is with the Department of Electrical and Computer Engineering, National University of Singapore (e-mail: elezhang@nus.edu.sg). The work of R. Zhang is supported by National University of Singapore under research grant R-261-518-005-720 and R-263-000-E86-112.}
			\author{\IEEEauthorblockN{Shuowen~Zhang, \emph{Member, IEEE} and Rui~Zhang, \emph{Fellow, IEEE}}}}
		\maketitle
		
		\begin{abstract}
			Intelligent reflecting surface (IRS) is a new promising technology that is able to reconfigure the wireless propagation channel via smart and passive signal reflection. In this paper, we investigate the \emph{capacity region} of a two-user communication network with one access point (AP) aided by $M$ IRS elements for enhancing the user-AP channels, where the IRS incurs negligible delay, thus the user-AP channels via the IRS follow the classic discrete memoryless channel model. In particular, we consider two practical IRS deployment strategies that lead to different effective channels between the users and AP, namely, the \emph{distributed deployment} where the $M$ elements form two IRSs, each deployed in the vicinity of one user, versus the \emph{centralized deployment} where all the $M$ elements are deployed in the vicinity of the AP. First, we consider the uplink multiple-access channel (MAC) and derive the capacity/achievable rate regions for both deployment strategies under different multiple access schemes. It is shown that the centralized deployment generally outperforms the distributed deployment under symmetric channel setups in terms of achievable user rates. Next, we extend the results to the downlink broadcast channel (BC) by leveraging the celebrated uplink-downlink (or MAC-BC) duality framework, and show that the superior rate performance of centralized over distributed deployment also holds. Numerical results are presented that validate our analysis, and reveal new and useful insights for optimal IRS deployment in wireless networks. 
		\end{abstract}
		
		\vspace{-5mm}
		\section{Introduction}\label{sec_intro}
		\vspace{-1mm}
		Driven by the recent advancement in metamaterial technology, \emph{intelligent reflecting surface (IRS)} has become a new solution to meet the increasingly high communication demands in beyond the fifth-generation (5G) wireless \hbox{networks} \cite{Survey,Towards,Survey_Basar}. Specifically, IRS is a reconfigurable metasurface consisting of a large number of passive elements, each of which is able to introduce an independent and controllable phase shift to the impinging electromagnetic wave, thereby collaboratively altering the propagation channels between the wireless transceivers. By properly designing the IRS reflection coefficients (i.e., phase shifts), IRS has been shown effective in adaptively reconfiguring the wireless environment for enhancing desired signal strength and mitigating interference, thus improving the achievable rate and/or reliability of various wireless communication systems. Moreover, IRSs are spectral-efficient and energy conservative since they perform passive signal reflection in the full-duplex mode without the need of any transmit radio frequency (RF) chains, which makes them suitable to be densely deployed in wireless networks.
		
		To fully reap the benefits of IRS, IRS needs to be efficiently integrated into existing wireless communication systems, which brings new challenges. First, it is of paramount importance to optimally design the IRS reflection coefficients such that the wireless channels are properly altered in favor of communication performance, which has attracted a great deal of research interests recently (see, e.g., \cite{MIMO_ISIT,MIMO,Protocol,SER,Emil,Secrecy_Zhang} for single-user systems and \cite{TWC_Yuen,Joint_Active,Discrete,WSR_Guo,OFDMA,random,Multicell,SWIPT,Robust_XH,NOMA_YW,Deployment_YW,FD,Joint,multicast} for multi-user systems). Particularly, from an information theoretical viewpoint, it is crucial to characterize the fundamental \emph{capacity limits} of IRS-aided communication systems so as to unveil the maximum performance gains brought by IRS, which, however, has only been pursued recently in \cite{MIMO_ISIT,MIMO,Beyond} under the single-user setup. While for the more complex IRS-aided multi-user systems of which the performance limit is characterized by the \emph{capacity region} that constitutes all the achievable rate-tuples of the users in the system, there has been very limited work in the literature, to the authors' best knowledge. This thus motivates the current work to characterize the capacity region of two fundamental multi-user communication systems aided by IRS, namely, the \emph{multiple-access channel (MAC)} in the uplink and the \emph{broadcast channel (BC)} in the downlink, respectively.
		
		\begin{figure*}[t]
			\centering
			\subfigure[Distributed deployment]{
				\includegraphics[height=5cm]{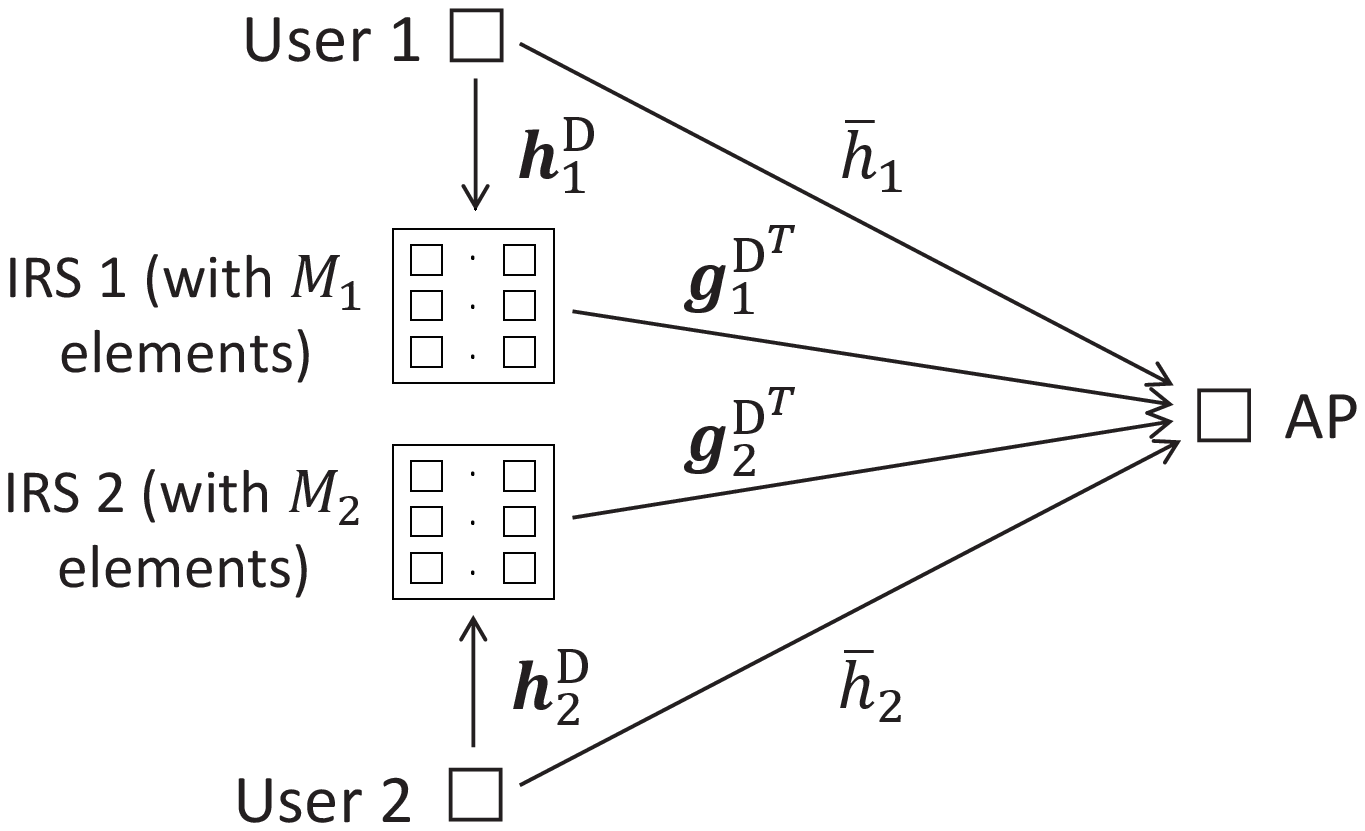}}
			\subfigure[Centralized deployment]{
				\includegraphics[height=5cm]{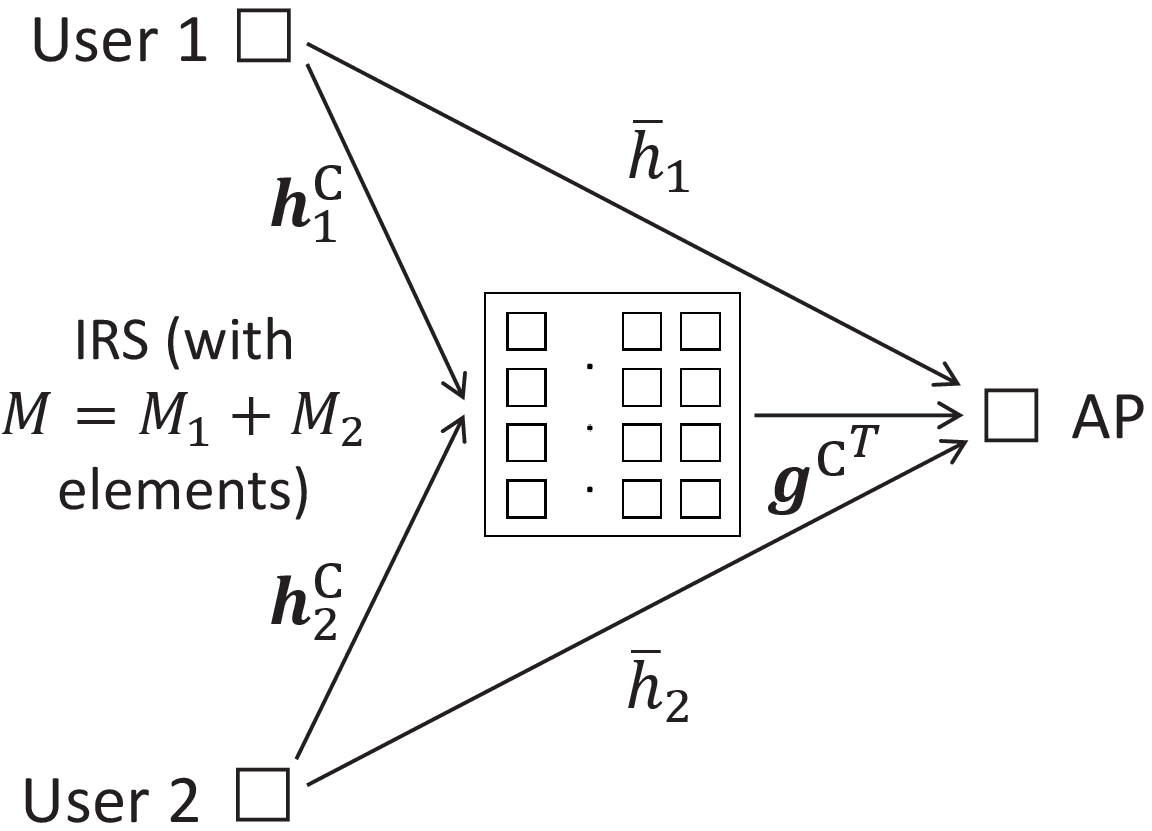}}
			\caption{An IRS-aided two-user communication system with different IRS deployment strategies (in the uplink MAC case).}\label{fig_system}
			
			\vspace{-5mm}
		\end{figure*}
		
		Besides capacity characterization through IRS reflection optimization, another important problem not well understood for IRS-aided multi-user systems is \emph{IRS deployment}. The existing works on IRS have mostly assumed \emph{given} IRS deployment (or IRS locations) without exploiting its design flexibility. However, with a given number of IRS elements, there are assorted approaches to deploy them in the network (e.g., by forming them as multiple IRSs, and placing them at different locations), which can lead to different IRS channels in general and thus impact the system capacity/achievable rates significantly. For instance, for the basic point-to-point communication system with one access point (AP) serving one user, if all elements form one single IRS, the IRS should be placed in close vicinity of the AP or the user so as to minimize the overall path loss of the AP-IRS-user channel, which increases with the product of the AP-IRS and IRS-user distances \cite{Survey}. In contrast, by splitting the elements into two IRSs and deploying them near the AP and the user, respectively, the received signal-to-noise ratio (SNR) can be significantly improved as compared to the optimal single-IRS deployment by exploiting the cooperative beamforming design of double IRSs, provided that the total number of IRS elements is sufficiently large \cite{Double_IRS}. However, for the more general system with multiple users or user clusters that are located far apart from one another, the optimal IRS deployment design has not been investigated yet. Generally speaking, motivated by the above single-user setup, there are two strategies to deploy a given number of IRS elements in the multi-user case: \emph{distributed IRS deployment} where the available elements form multiple distributed IRSs each deployed near one user (or user cluster), as illustrated in Fig. \ref{fig_system} (a), and \emph{centralized IRS deployment} where all elements form one IRS deployed near the AP, as illustrated in Fig. \ref{fig_system} (b). Note that these two strategies will lead to different user-AP effective channels in general and hence different user achievable rates. Specifically, with distributed deployment, each user can only enjoy the passive beamforming gain brought by its nearby IRS (since its signals reflected by other far-apart IRSs are too weak due to much higher path loss), which is thus smaller than the passive beamforming gain under the centralized deployment with a larger-size IRS where all the elements can be used for enhancing the channel of any user. However, the IRS passive beamforming gain under the centralized deployment needs to be shared by all the simultaneously served users in general, which may result in a reduced gain for each user. To the best of our knowledge, it is yet unclear which of the above two IRS deployment strategies achieves larger capacity region or achievable rate regions with practical orthogonal multiple access (OMA) schemes such as time-division multiple access (TDMA) and frequency-division multiple access (FDMA) in multi-user systems such as MAC and BC.
		
		To tackle the above capacity characterization and IRS deployment problems, we study in this paper a two-user communication system aided by $M$ IRS elements, as shown in Fig. \ref{fig_system}. Specifically, this paper considers the most practical scenario of employing IRS, where the phase shifts at the IRS elements are tuned to collaboratively alter the effective user-AP channels only. We aim to derive the capacity region as well as achievable rate regions with TDMA/FDMA under the proposed two IRS deployment strategies, for both the uplink MAC and downlink BC. To pursue the fundamental performance limit, we assume the availability of perfect channel state information (CSI) for all the channels shown in Fig. \ref{fig_system}, which can be acquired via various existing channel training and estimation methods proposed in e.g., \cite{CE_B,CE_Liu,CE_Johansson,CE_Yuan,Protocol,CE_You,CE_Zheng}. Our main contributions are summarized as follows.
		\begin{itemize}[leftmargin=*]
			\item First, we investigate the uplink MAC. For the distributed IRS deployment, we provide \emph{closed-form} characterizations of its capacity region as well as the achievable rate regions with TDMA and FDMA. For the centralized IRS deployment, we develop a \emph{rate-profile} based method to characterize the capacity region by solving a series of sum-rate maximization problems via joint IRS reflection and transmit power optimization, each corresponding to a different rate-ratio between the two users. An efficient alternating optimization (AO) algorithm is proposed to find a high-quality approximate solution to each problem, where we iteratively obtain the \emph{optimal} solution to each IRS reflection coefficient with the others being fixed. Based on this, a capacity region inner bound (or achievable rate region) can be obtained with polynomial complexity. We further propose a capacity region outer bound via the semi-definite relaxation (SDR) technique. In addition, we characterize the achievable rate region with TDMA in closed-form and that with FDMA via a similar rate-profile method.
			\item Moreover, we analytically prove that the capacity region as well as achievable rate regions with TDMA and FDMA for the case of centralized deployment \emph{contain} the corresponding regions for the case of distributed deployment under a practical symmetric channel setup. Furthermore, despite the lack of frequency-selective signal reflection at the IRS, the achievable rate region with FDMA still contains that with TDMA for both deployment strategies.
			\item Next, we extend the results for the uplink MAC to the downlink BC by leveraging the \emph{uplink-downlink (or MAC-BC) duality} framework. We propose computationally efficient methods to characterize the capacity region and achievable rate regions with TDMA/FDMA based on those for the dual MAC, and prove that the performance advantage of centralized over distributed IRS deployment is also valid for the downlink BC.
			\item Finally, numerical results validate our analysis as well as tightness of the proposed bounds. By comparing the various capacity/rate regions, we also draw useful insights into  optimal deployment of IRSs in practical systems. Particularly, it is shown that the capacity gain of centralized over distributed IRS deployment is most prominent when the rates of the two users are \emph{asymmetric}; moreover, centralized IRS deployment is more effective in alleviating the \emph{``near-far''} problem in multi-user communications, as a result of users with drastically different distances from the AP.
		\end{itemize}
		
		The rest of this paper is organized as follows. Section \ref{sec_sys} presents the system models under the two IRS deployment strategies. For the uplink MAC, Section \ref{sec_dis} and Section \ref{sec_cen} characterize the capacity region and achievable rate regions with TDMA/FDMA under distributed and centralized IRS deployment, respectively; Section \ref{sec_comparison} compares these regions under the two IRS deployment strategies. Section \ref{sec_BC} extends the above results to the downlink BC. Numerical examples and their pertinent discussions are presented in Section \ref{sec_num}. Finally, Section \ref{sec_con} concludes this paper.
		
		\emph{Notations:} Vectors and matrices are denoted by boldface lower-case letters and boldface upper-case letters, respectively. $|x|$, $x^*$, $\arg\{x\}$, and $\mathfrak{Re}\{x\}$ denote the absolute value, conjugate, angle, and real part of a complex number $x$, respectively. For a complex vector $\mv{z}$, $\|{\mv{z}}\|_p$ and $z_k$ denote the $l_p$-norm and the $k$th element, respectively, and $\mathrm{diag}\{{\mv{z}}\}$ denotes a square diagonal matrix with the elements of $\mv{z}$ on its main diagonal. $\mathbb{C}^{M\times N}$ denotes the space of $M\times N$ complex matrices. ${\mv{0}}$ denotes an all-zero matrix with appropriate dimension. For an $M\times N$ matrix $\mv{A}$, $\mathrm{rank}({\mv{A}})$ and $[{\mv{A}}]_{i,j}$ denote the rank and $(i,j)$-th element of $\mv{A}$, respectively. For a square matrix $\mv{S}$, $\mv{S}\succeq {\mv{0}}$ means that $\mv{S}$ is positive semi-definite. The distribution of a circularly symmetric complex Gaussian (CSCG) random variable with mean $0$ and variance $\sigma^2$ is denoted by $\mathcal{CN}(0,\sigma^2)$; and $\sim$ stands for ``distributed as''. $\mathbb{E}[\cdot]$ denotes the statistical expectation. $\mathcal{O}(\cdot)$ denotes the standard big-O notation. $\mathrm{Conv}(\cdot)$ denotes the convex hull operation. $\bigcup$ denotes the union operation.
		
		\vspace{-4mm}
		\section{System Model}\label{sec_sys}
		\vspace{-1mm}
		We consider a communication network where one single-antenna AP serves two single-antenna users that are sufficiently far apart from each other, as illustrated in Fig. \ref{fig_system}. 
		In the uplink, each user aims to send an independent message to the AP; while in the downlink, the AP sends independent messages to the users. Moreover, $M$ passive IRS elements are deployed to enhance the user communication rates. Specifically, this paper considers the most practical scenario of employing IRS, where each IRS element induces an independent phase shift to its incident signal for collaboratively altering the effective channels between the users and the AP.\footnote{It is worth noting that IRS can perform other functions besides altering the wireless channel. For example, if user messages are known at the IRS, IRS can forward messages to the users by varying its reflection over time \cite{Beyond}. However, this requires message sharing between the AP and IRS, which is not considered in this paper.} Under this strategy, we define the \emph{IRS-aided MAC} and the \emph{IRS-aided BC} as the uplink (from the users to the AP) and downlink (from the AP to the users) transmissions, respectively, which are similar to the conventional discrete memoryless MAC and BC \cite{Elements}, but with the user effective channels controllable by the IRS.
		
		For the IRS-aided MAC and BC, we propose two different deployment strategies for the $M$ elements. Specifically, for the \emph{distributed deployment}, the $M$ elements form two IRSs (see Fig. \ref{fig_system} (a)), where IRS $k$, $k\in \{1,2\}$, consists of $M_k$ elements, with $M_k\geq 1$, and is placed in the vicinity of user $k$, subject to $\sum_{k=1}^2 M_k=M$. In contrast, for the \emph{centralized deployment}, all the $M$ elements form one single IRS located in the vicinity of the AP (see Fig. \ref{fig_system} (b)). For the purpose of exposition, we will first focus our study on the IRS-aided MAC in the uplink shown in Fig. \ref{fig_system}, and then extend our study to the IRS-aided BC in the downlink (see Section \ref{sec_BC}). Specifically, for the IRS-aided MAC, we denote the baseband equivalent direct channel from user $k$ to the AP as $\bar{h}_k\in \mathbb{C}$, $k=1,2$. In the following, we describe the system models for the two IRS deployment cases, respectively.
		
		\vspace{-3mm}
		\subsection{Distributed IRS Deployment}\label{sec_sys_dis}
		\vspace{-1mm}
		For distributed IRS deployment, we denote ${\mv{h}}_k^\dd\in \mathbb{C}^{M_k\times 1}$ as the channel vector from user $k$ to its serving (nearby) IRS, and ${\mv{g}}_k^{\dd^T}\in \mathbb{C}^{1\times M_k}$ as the channel vector from its serving IRS to the AP. Denote ${\mv{\Phi}}^\dd_k=\mathrm{diag}\{\phi^\dd_{k1},...,\phi^\dd_{k M_k}\}\in \mathbb{C}^{M_k\times M_k}$ as the IRS reflection matrix for IRS $k$, with $|\phi^\dd_{km}|=1,\ \forall m\in {\mathcal{M}}_k$, where ${\mathcal{M}}_k=\left\{1,...,M_k\right\}$. Since the locations of the two users are sufficiently far apart, we assume that the signal transmitted by one user and reflected by the other user's serving IRS is negligible at the AP due to the severe path loss. Hence, the effective channel from user $k$ to the AP by combining the direct link and the reflected link by its serving IRS is given by
		\begin{equation}\label{channel_distributed}
			\tilde{h}^\dd_k(\mv{\Phi}_k^\dd)=\bar{h}_k+{\mv{g}}_k^{\dd^T}{\mv{\Phi}}^\dd_k{\mv{h}}_k^\dd,\quad k=1,2.
		\end{equation}
		Let $s_k\in \mathbb{C}$ denote the desired information symbol for user $k$ with zero mean and unit variance. Note that $s_k$'s are independent over $k$. The transmitted signal by user $k$ is modeled as $x_k=\sqrt{p_k}s_k$, which satisfies $\mathbb{E}[|x_k|^2]=p_k\leq P_k$, with $p_k$ denoting the transmit power of user $k$ and $P_k$ denoting its maximum value. The received signal at the AP is thus modeled as
		\begin{equation}\label{received_distributed}
			y=\tilde{h}^\dd_1(\mv{\Phi}_1^\dd)x_1+\tilde{h}^\dd_2(\mv{\Phi}_2^\dd)x_2+z,
		\end{equation}
		where $z\sim \mathcal{CN}(0,\sigma^2)$ denotes the CSCG noise at the AP receiver with average power $\sigma^2$. For each user $k$, let $R^\dd_k$ denote its achievable rate in bits per second per Hertz (bps/Hz) under the distributed IRS deployment.
		
		\vspace{-3mm}
		\subsection{Centralized IRS Deployment}\label{set_sys_cen}
		\vspace{-1mm}
		For centralized IRS deployment, we denote $\mv{h}_k^\cc\in \mathbb{C}^{M\times 1}$ as the channel vector from user $k$ to the (single) IRS, and $\mv{g}^{\cc^T}\in \mathbb{C}^{1\times M}$ as the channel vector from the IRS to the AP. Denote $\mv{\Phi}^\cc=\mathrm{diag}\{\phi_1^\cc,...,\phi_M^\cc\}\in \mathbb{C}^{M\times M}$ as the IRS reflection matrix, with $|\phi^\cc_m|=1,\ \forall m\in \mathcal{M}$, where $\mathcal{M}=\{1,...,M\}$. Thus, the effective channel from user $k$ to the AP is given by
		\begin{equation}\label{channel_centralized}
			\tilde{h}^\cc_k(\mv{\Phi}^\cc)=\bar{h}_k+\mv{g}^{\cc^T}\mv{\Phi}^\cc\mv{h}_k^\cc,\quad k=1,2.
		\end{equation}
		Note that different from the distributed deployment where the effective channel between user $k$ and the AP is only dependent on the $M_k$ reflection coefficients of its own serving IRS in $\mv{\Phi}_k^\dd$, the effective channels for both users under the centralized deployment depend on all the $M$ reflection coefficients in $\mv{\Phi}^\cc$. Under the same transmitted signal and receiver noise model as in the distributed deployment case, the received signal at the AP is modeled similarly as (\ref{received_distributed}) by replacing each $\tilde{h}^\dd_k(\mv{\Phi}_k^\dd)$ with $\tilde{h}^\cc_k(\mv{\Phi}^\cc)$. For user $k$, let $R^\cc_k$ denote its achievable rate in bps/Hz under the centralized IRS deployment.
		
		In the following, we characterize the \emph{capacity region} of the IRS-aided two-user MAC under each of the two deployment strategies, which constitutes all the achievable rate-pairs $(R^\dd_1,R^\dd_2)$'s or $(R^\cc_1,R^\cc_2)$'s. We also derive their \emph{achievable rate regions} under practical OMA schemes including TDMA and FDMA, where the two users communicate with the AP in orthogonal time slots or frequency bands, respectively. We then compare these capacity (rate) regions and draw useful insights into the optimal IRS deployment strategy.
		
		\vspace{-3mm}
		\section{Distributed IRS Deployment}\label{sec_dis}
		\vspace{-1mm}
		In this section, we characterize the capacity region as well as the achievable rate regions with TDMA/FDMA under the distributed IRS deployment.
		
		\vspace{-3mm}
		\subsection{Capacity Region}\label{sec_dis_capacity}
		\vspace{-1mm}
		First, we derive the capacity region to unveil the fundamental limit. Note that with given IRS reflection coefficients $\{{\mv{\Phi}}^\dd_k\}$, the channels from the two users to the AP are determined as $\{\tilde{h}_k^\dd(\mv{\Phi}_k^\dd)\}$ given in (\ref{channel_distributed}), and the capacity region of the two-user MAC is well-known as the pentagon region consisting of all rate-pairs that satisfy the following constraints \cite{Elements}:
		\begin{align}
			{R}_1^\dd\leq &\log_2(1+{P_1|{\tilde{h}}^\dd_1(\mv{\Phi}_1^\dd)|^2}/{\sigma^2})\overset{\Delta}{=}r^\dd_1({\mv{\Phi}}^\dd_1),\label{D1}\\
			{R}_2^\dd\leq &\log_2(1+{P_2|{\tilde{h}}^\dd_2(\mv{\Phi}_2^\dd)|^2}/{\sigma^2})\overset{\Delta}{=}r^\dd_2({\mv{\Phi}}^\dd_2),\label{D2}\\
			{R}_1^\dd+{R}_2^\dd\leq &\log_2(1+(P_1|{\tilde{h}}^\dd_1(\mv{\Phi}_1^\dd)|^2+P_2|{\tilde{h}}^\dd_2(\mv{\Phi}_2^\dd)|^2)/\sigma^2)\nonumber\\
			&\  \overset{\Delta}{=}r^\dd_{12}(\{{\mv{\Phi}}^\dd_k\}),\label{D3}
		\end{align}
		which is denoted as ${\mathcal{C}}^\dd(\{{\mv{\Phi}}^\dd_k\})$. Note that by flexibly designing the IRS reflection coefficients $\{\mv{\Phi}^\dd_k\}$, any rate-pair within the union set of ${\mathcal{C}}^\dd(\{{\mv{\Phi}}^\dd_k\})$'s over all feasible $\{\mv{\Phi}^\dd_k\}$'s can be achieved. By further considering \emph{time sharing} among different $\{\mv{\Phi}^\dd_k\}$'s, the capacity region of IRS-aided MAC for distributed IRS deployment is defined as the convex hull of such a union set \cite{Elements}:
		\begin{equation}\label{Cdis}
			\mathcal{C}^\dd\overset{\Delta}{=}\mathrm{Conv}\Big({\bigcup}_{\{{\mv{\Phi}}^\dd_k\}\in \mathcal{F}^\dd} {\mathcal{C}}^\dd(\{{\mv{\Phi}}^\dd_k\})\Big),
		\end{equation}
		where $\mathcal{F}^\dd\overset{\Delta}{=}\{{\{{\mv{\Phi}}^\dd_k\}:|\phi^\dd_{km}|=1,\forall k, m}\}$ denotes the feasible set of ${\{{\mv{\Phi}}^\dd_k\}}$. It is worth noting that for any given $\{{\mv{\Phi}}^\dd_k\}$, the optimal input distribution for achieving ${\mathcal{C}}^\dd(\{{\mv{\Phi}}^\dd_k\})$ is the CSCG distribution with $s_k\sim \mathcal{CN}(0,1),\ \forall k$ \cite{Elements}. Therefore, it follows from (\ref{Cdis}) that the capacity-achieving optimal input distribution for the IRS-aided MAC under distributed deployment is still $s_k\sim \mathcal{CN}(0,1),\ \forall k$, which will be considered throughout this section.
		
		In the following, we characterize $\mathcal{C}^\dd$ in closed-form by exploiting the peculiar effective channel structure under the distributed deployment. Specifically, according to the triangle inequality, for any $\{\mv{\Phi}_k^\dd\}\in\mathcal{F}^\dd$, the effective channel gain for each user $k$ is upper-bounded by
		\begin{align}\label{eq1}
			|\tilde{h}_k^\dd(\mv{\Phi}_k^\dd)|=&|\bar{h}_k+{\textstyle\sum}_{m=1}^{M_k}{g}^\dd_{km}{\phi}^\dd_{km}{h}_{km}^\dd|\nonumber\\
			\leq&|\bar{h}_k|+{\textstyle\sum}_{m=1}^{M_k}|g^\dd_{km}||h_{km}^\dd|\overset{\Delta}{=}\tilde{h}_{k,\ub}^\dd,\ k=1,2,
		\end{align}
		where the inequality holds with equality if and only if $\{\mv{\Phi}_k^\dd\}$ is designed as follows:
		\begin{equation}\label{D_phi}
			\phi_{km}^\dd=e^{j(\arg\{\bar{h}_k\}-\arg\{g^\dd_{km}h_{km}^\dd\})},\quad k=1,2,\ m\in \mathcal{M}_k.
		\end{equation}
		Based on this result, we obtain the following theorem.
		\begin{theorem}\label{theorem_dis}
		The capacity region of the IRS-aided two-user MAC in (\ref{channel_distributed}) under the distributed deployment is given by
		\begin{equation}\label{C_dis}
				\mathcal{C}^\dd=\{(R^\dd_1,R^\dd_2):\!R^\dd_1\leq r_1^{\dd^\star},R^\dd_2\leq r_2^{\dd^\star},R^\dd_1+R^\dd_2\leq r_{12}^{\dd^\star}\},
		\end{equation}
			where $r_1^{\dd^\star}\overset{\Delta}{=}\log_2(1\!+\!{P_1\tilde{h}_{1,\ub}^{\dd^2}}/{\sigma^2})$, $r_2^{\dd^\star}\overset{\Delta}{=}\log_2(1+{P_2\tilde{h}_{2,\ub}^{\dd^2}}/{\sigma^2})$, and $r_{12}^{\dd^\star}\overset{\Delta}{=}\log_2\big(1+(P_1\tilde{h}_{1,\ub}^{\dd^2}+P_2\tilde{h}_{2,\ub}^{\dd^2})/\sigma^2\big)$.
		\end{theorem}
		\begin{IEEEproof}
			Theorem \ref{theorem_dis} can be proved by noting that $\mathcal{C}^\dd$ given in (\ref{C_dis}) is an achievable rate region with $\{\mv{\Phi}_k^\dd\}$ given in (\ref{D_phi}), and also provides a convex-shape outer bound for all achievable ${\mathcal{C}}^\dd(\{{\mv{\Phi}}^\dd_k\})$'s given in (\ref{D1})--(\ref{D3}) (thus, the convex-hull operation in (\ref{Cdis}) is not needed with $\{{\mv{\Phi}}^\dd_k\}$ given in (\ref{D_phi})).
		\end{IEEEproof}
		
		Note that to achieve the above capacity region $\mathcal{C}^\dd$, successive interference cancellation (SIC) or joint decoding needs to be performed at the AP in general \cite{Elements}. For example, with SIC, the AP needs to first decode the message of one user by treating the signal of the other user as noise, then cancel the decoded signal and decode the other user's message \cite{Elements}. Next, we derive the achievable rate regions with TDMA and FDMA, where the SIC/joint decoding operation is not needed since the signals of the two users are already separated in the time and frequency domains, respectively. Note that these achievable rate regions generally serve as inner bounds of the capacity region.
		
		\vspace{-3mm}
		\subsection{Achievable Rate Region with TDMA}\label{sec_dis_TDMA}
		\vspace{-1mm}
		With TDMA, the two users transmit in two orthogonal time slots, where we let $\rho_{\ttt}\in [0,1]$ denote the fraction of time that user $1$ sends its message. In this case, under distributed IRS deployment, the achievable rate of each user $k$ only depends on the reflection matrix of IRS $k$ at its assigned time slot, which is denoted as $\mv{\Phi}_k^{\dd}[k]$. For any given $\{\mv{\Phi}_k^{\dd}[k]\}$, the achievable rate region is defined as  $\mathcal{R}^{\dd}_{\ttt}(\{\mv{\Phi}_k^{\dd}[k]\})={\bigcup}_{\rho_{\ttt}\in [0,1]}\{(R_1^\dd,R_2^\dd):
		R_1^\dd\leq \rho_{\ttt}\log_2(1+P_1|\tilde{h}_1^\dd(\mv{\Phi}_1^\dd[1])|^2/\sigma^2),
		R_2^\dd\leq (1-\rho_{\ttt})\log_2(1+P_2|\tilde{h}_2^\dd(\mv{\Phi}_2^\dd[2])|^2/\sigma^2)\}$. Note that both $|\tilde{h}_1^\dd(\mv{\Phi}_1^\dd[1])|$ and $|\tilde{h}_2^\dd(\mv{\Phi}_2^\dd[2])|$ can be maximized as $\tilde{h}_{1,\ub}^\dd$ and $\tilde{h}_{2,\ub}^\dd$ in (\ref{eq1}) by setting $\mv{\Phi}_1^{\dd}[1]$ and $\mv{\Phi}_2^{\dd}[2]$ as $\mv{\Phi}_1^{\dd}$ and $\mv{\Phi}_2^{\dd}$ given in (\ref{D_phi}), respectively; moreover, $\mathcal{R}^{\dd}_{\ttt}(\{\mv{\Phi}_k[k]\})$ with the aforementioned $\{\mv{\Phi}_k[k]\}$ can be easily shown to be a convex region. Therefore, it can be proved similarly as Theorem \ref{theorem_dis} that the achievable rate region with TDMA under distributed deployment is
		\begin{align}
			\mathcal{R}^{\dd}_{\ttt}=&{\bigcup}_{\rho_{\ttt}\in [0,1]}\Bigg\{(R_1^\dd,R_2^\dd):
			R_1^\dd\leq \rho_{\ttt}\log_2\left(1+\frac{P_1\tilde{h}_{1,\ub}^{\dd^2}}{\sigma^2}\right),\nonumber\\
			&R_2^\dd\leq (1-\rho_{\ttt})\log_2\left(1+\frac{P_2\tilde{h}_{2,\ub}^{\dd^2}}{\sigma^2}\right)\Bigg\}.\label{T_D}
		\end{align} 
	
		\vspace{-3mm}
		\subsection{Achievable Rate Region with FDMA}\label{sec_dis_FDMA}
		\vspace{-1mm}
		With FDMA, the two users transmit simultaneously over two orthogonal frequency bands, where we let $\rho_{\ff}\in [0,1]$ denote the fraction of bandwidth assigned to user $1$.\footnote{It is worth noting that the IRS reflection coefficients impact the channels at different frequency bands identically without frequency selectivity, due to the lack of RF chains and baseband processing at the IRS.} With any given $\{\mv{\Phi}_k^{\dd}\}$, the achievable rate region is given by $\mathcal{R}^{\dd}_{\ff}(\{\mv{\Phi}_k^{\dd}\})={\bigcup}_{\rho_{\ff}\in [0,1]}\{(R_1^\dd,R_2^\dd):R_1^\dd\leq \rho_{\ff}\log_2(1+P_1|\tilde{h}_1^\dd(\mv{\Phi}_1^\dd)|^2/(\rho_{\ff}\sigma^2)),R_2^\dd\leq (1-\rho_{\ff})\log_2(1+P_2|\tilde{h}_2^\dd(\mv{\Phi}_2^\dd)|^2/((1-\rho_{\ff})\sigma^2))\}$ \cite{Elements}. Note that $|\tilde{h}_1^\dd(\mv{\Phi}_1^\dd)|$ and $|\tilde{h}_2^\dd(\mv{\Phi}_2^\dd)|$ can be simultaneously maximized as $\tilde{h}_{1,\ub}^\dd$ and $\tilde{h}_{2,\ub}^\dd$ in (\ref{eq1}) with $\{\mv{\Phi}_k^{\dd}\}$ given in (\ref{D_phi}), and the corresponding $\mathcal{R}^{\dd}_{\ff}(\{\mv{\Phi}_k\})$ is a convex region \cite{Elements}. Thus, similar to the TDMA case, the achievable rate region with FDMA under distributed deployment is
		\begin{align}
			\mathcal{R}^{\dd}_{\ff}=&{\bigcup}_{\rho_{\ff}\in [0,1]}\Bigg\{(R_1^\dd,R_2^\dd):
			R_1^\dd\leq \rho_{\ff}\log_2\left(1+\frac{P_1\tilde{h}_{1,\ub}^{\dd^2}}{\rho_{\ff}\sigma^2}\right),\nonumber\\
			&R_2^\dd\leq (1-\rho_{\ff})\log_2\left(1+\frac{P_2\tilde{h}_{2,\ub}^{\dd^2}}{(1-\rho_{\ff})\sigma^2}\right)\Bigg\}.\label{T_F}
		\end{align}
		
		To summarize, the Pareto boundaries of all the capacity region and TDMA/FDMA achievable rate regions under distributed IRS deployment are achieved by setting the reflection coefficients at each IRS based on its nearby user's channel as given in (\ref{D_phi}). Moreover, by comparing (\ref{T_F}) with (\ref{T_D}), it can be easily shown that the achievable rate region of FDMA contains that of TDMA since in both cases the optimal reflection coefficients are identical and thus yield the same user-AP effective channels, while the signal energy transmitted by the two users in TDMA is generally less than their counterparts in FDMA \cite{Elements}. In addition, $\mathcal{R}^{\dd}_{\ff}\subseteq \mathcal{C}^\dd$ holds since $\mathcal{R}^{\dd}_{\ff}(\{\mv{\Phi}_k^\dd\})\subseteq \mathcal{C}^\dd(\{\mv{\Phi}_k^\dd\})$ holds for any $\{\mv{\Phi}_k^\dd\}$ \cite{Elements}, and the optimal $\{\mv{\Phi}_k^\dd\}$'s that achieve the Pareto boundaries of $\mathcal{C}^\dd$ and $\mathcal{R}^{\dd}_{\ff}$ are the same, as shown in Section \ref{sec_dis_capacity} and Section \ref{sec_dis_FDMA}. Therefore, we have $\mathcal{R}^{\dd}_{\ttt}\subseteq \mathcal{R}^{\dd}_{\ff}\subseteq \mathcal{C}^\dd$ in the case of distributed IRS deployment.
		
		\vspace{-3mm}
		\section{Centralized IRS Deployment}\label{sec_cen}
		\vspace{-1mm}
		In this section, we characterize the capacity region and achievable rate regions with TDMA/FDMA under the centralized IRS deployment. Similar to the distributed IRS case, the capacity region with given reflection coefficients ${\mv{\Phi}}^\cc$ of the centralized IRS is the pentagon region consisting of all rate-pairs that satisfy the following constraints:
		\begin{align}
			{R}_1^\cc\leq &\log_2(1+P_1|{\tilde{h}}^\cc_1(\mv{\Phi}^\cc)|^2/\sigma^2)\overset{\Delta}{=}r^\cc_1(\mv{\Phi}^\cc),\label{C1}\\
			{R}_2^\cc\leq &\log_2(1+P_2|{\tilde{h}}^\cc_2(\mv{\Phi}^\cc)|^2/\sigma^2)\overset{\Delta}{=}r^\cc_2(\mv{\Phi}^\cc),\label{C2}\\
			{R}_1^\cc+{R}_2^\cc\leq& \log_2(1+(P_1|{\tilde{h}}^\cc_1({\mv{\Phi}}^\cc)|^2+P_2|{\tilde{h}}^\cc_2({\mv{\Phi}}^\cc)|^2)/\sigma^2)\nonumber\\
			& \ \overset{\Delta}{=}r^\cc_{12}(\mv{\Phi}^\cc),\label{C3}
		\end{align}
		which is denoted as ${\mathcal{C}}^\cc({\mv{\Phi}}^\cc)$. By tuning the IRS reflection coefficients $\mv{\Phi}^\cc$ and performing time sharing among different $\mv{\Phi}^\cc$'s, the capacity region is defined as
		\begin{equation}\label{C_cen}
			\mathcal{C}^\cc\overset{\Delta}{=}\mathrm{Conv}\Big({\bigcup}_{{\mv{\Phi}}^\cc\in \mathcal{F}^\cc} {\mathcal{C}}^\cc({\mv{\Phi}}^\cc)\Big),
		\end{equation}
		where $\mathcal{F}^\cc\overset{\Delta}{=}\{{\mv{\Phi}}^\cc:|\phi^\cc_{m}|=1,\forall m\}$ denotes the feasible set of ${\mv{\Phi}}^\cc$. Similar to the distributed case, for any given ${\mv{\Phi}}^\cc$, the optimal input distribution that achieves the capacity region ${\mathcal{C}}^\cc({\mv{\Phi}}^\cc)$ is $s_k\sim \mathcal{CN}(0,1),\ \forall k$. It thus follows from (\ref{C_cen}) that it is also the capacity-achieving input distribution for the IRS-aided MAC under centralized deployment, which will be assumed throughout this section.
		
\begin{figure}[t]
			\centering
			\includegraphics[width=7cm]{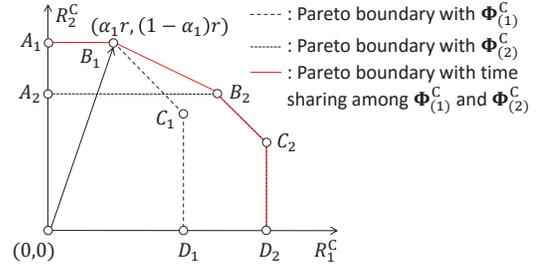}
			\caption{Illustration of the capacity region under centralized IRS deployment. SIC-achievable rate-pairs with given IRS reflection coefficients $\mv{\Phi}_{(n)}^\cc$ are shown on the line segments $A_n$--$B_n$ and $C_n$--$D_n$, $n=1,2$.}\label{rateprofile}
		\end{figure}

		Compared to the distributed IRS case, the capacity region in (\ref{C_cen}) is more challenging to characterize. This is because the effective channels of the two users, $\tilde{h}_1^\cc(\mv{\Phi}^\cc)$ and $\tilde{h}_2^\cc(\mv{\Phi}^\cc)$, are \emph{coupled} through all the $M$ reflection coefficients in $\mv{\Phi}^\cc$. Thus, different portions of the Pareto boundary of the capacity region $\mathcal{C}^\cc$ are generally achieved by different $\mv{\Phi}^\cc$ to strike a balance between $\tilde{h}_1^\cc(\mv{\Phi}^\cc)$ and $\tilde{h}_2^\cc(\mv{\Phi}^\cc)$, as illustrated in Fig. \ref{rateprofile}. Finding such capacity-achieving sets of reflection coefficients is more challenging as compared to the distributed IRS case where the entire Pareto boundary of the capacity region is achieved by a single set of $\{\mv{\Phi}_k^\dd\}$ given in (\ref{D_phi}), since the effective channel of each user is maximized by the reflection coefficients of its own serving IRS.\footnote{It is worth noting that the user channels under distributed deployment can be expressed as a special case of those under centralized deployment, thus the capacity/rate region characterization methods for the centralized case are also readily applicable to the distributed case.} Although $\mathcal{C}^\cc$ can be characterized via the \emph{exhaustive search} method by first obtaining $\mathcal{C}^\cc(\mv{\Phi}^\cc)$'s for all feasible ${\mv{\Phi}}^\cc\in \mathcal{F}^\cc$ and then taking the convex hull of their union set, the required complexity is at least $\mathcal{O}(L_0^M)$ if the $[0,2\pi)$ phase range for each $\phi_m^\cc$ in ${\mv{\Phi}}^\cc$ is approximated by $L_0$ uniformly sampled points, which is \emph{exponential} over $M$ and thus prohibitive for practically large $M$. To avoid such high complexity for characterizing the capacity region $\mathcal{C}^\cc$, in the following subsections, we provide an alternative method to characterize $\mathcal{C}^\cc$, and develop efficient algorithms to find both inner and outer bounds of $\mathcal{C}^\cc$, whose tightness will be evaluated via numerical results in Section \ref{sec_num}.
		
		\vspace{-4mm}
		\subsection{Rate-Profile based Capacity Region Characterization}\label{sec_cen_cha}
		\vspace{-1mm}
		To start with, note that for each $\mv{\Phi}^\cc$, all the achievable rate-pairs on the Pareto boundary of its corresponding $\mathcal{C}^\cc(\mv{\Phi}^\cc)$ except those requiring time sharing/rate splitting of the two users can be attained via SIC at the AP \cite{Elements}. Motivated by this result, we propose to first characterize the Pareto boundary of the union set of the above SIC-achievable rate-pairs for all feasible $\mv{\Phi}^\cc\in \mathcal{F}^\cc$, and then perform time sharing among the obtained rate-pairs on the Pareto boundary to further enlarge the achievable rate region, as illustrated in Fig. \ref{rateprofile}. For the first task, we propose to adopt the \emph{rate-profile} approach in \cite{Cooperative}.\footnote{It is worth noting that another approach to characterize the Pareto boundary is by solving a series of \emph{weighted sum-rate maximization (WSRmax)} problems \cite{Cooperative}. However, this approach is generally not guaranteed to obtain the complete Pareto boundary $\mathcal{C}^\cc$ since the region formed by the SIC-achievable rate-pairs without time sharing is non-convex in general \cite{Cooperative}; moreover, such WSRmax problems are also challenging to solve since the rates of the two users are coupled in the objective function in a complicated manner.} Specifically, let $\mv{\pi}$ denote the decoding order indicator, with $\mv{\pi}=[1,2]^T\overset{\Delta}{=}\mv{\pi}^{\II}$ representing that user $1$ is decoded before user $2$, and $\mv{\pi}=[2,1]^T\overset{\Delta}{=}\mv{\pi}^{\II\II}$ otherwise. Let $\alpha_{\pi_1}\in [0,1]$ denote the rate ratio between the firstly decoded user and the users' sum-rate. We further denote $\alpha_1\in [0,1]$ as the rate ratio between user 1 and the users' sum-rate, $\alpha_2=1-\alpha_1\in [0,1]$ as that between user 2 and the users' sum-rate, and $\mv{\alpha}=[\alpha_1,1-\alpha_1]^T$ as the rate-profile vector, with $\mv{\alpha}=[\alpha_{\pi_1},1-\alpha_{\pi_1}]^T$ if $\mv{\pi}=\mv{\pi}^\II$ and $\mv{\alpha}=[1-\alpha_{\pi_1},\alpha_{\pi_1}]^T$ if $\mv{\pi}=\mv{\pi}^{\II\II}$. Based on the above, we formulate the following problem to maximize the sum-rate of the two users with given $\mv{\alpha}$ and $\mv{\pi}$ by jointly optimizing \hbox{the IRS reflection coefficients and user transmit powers:}
		\begin{align}
			\!\!\!\!\mbox{(P1)}\underset{r,p_1,p_2,\mv{\Phi}^\cc}{\mathtt{max}} &r\\
			\mathtt{s.t.}\  &\log_2\bigg(1+\frac{p_{\pi_1}|\tilde{h}_{\pi_1}^\cc(\mv{\Phi}^\cc)|^2}{p_{\pi_2}|\tilde{h}_{\pi_2}^\cc(\mv{\Phi}^\cc)|^2+\sigma^2}\bigg)\geq \alpha_{\pi_1} r\label{P1c1}\\
			&\log_2\left(1+\frac{p_{\pi_2}|\tilde{h}_{\pi_2}^\cc(\mv{\Phi}^\cc)|^2}{\sigma^2}\right)\geq (1\!-\!\alpha_{\pi_1}) r\label{P1c2}\\
			& p_k\leq P_k,\quad  \forall k\in\{1,2\}\label{P1c3}\\
			& \mv{\Phi}^\cc=\mathrm{diag}\{\phi_1^\cc,...,\phi_M^\cc\}\label{P1c4}\\
			& |\phi_m^\cc|=1,\quad \forall m\in \mathcal{M}.\label{P1c5}
		\end{align}
		
		For each rate-profile vector $\mv{\alpha}$, let $r_\II^\star(\mv{\alpha})$ and $r_{\II\II}^\star(\mv{\alpha})$ denote the optimal value to (P1) with $\mv{\pi}=\mv{\pi}^{\II}$ and $\mv{\pi}=\mv{\pi}^{\II\II}$, respectively. Note that $r_\II^\star(\mv{\alpha})\geq r_{\II\II}^\star(\mv{\alpha})$ represents that decoding order $\mv{\pi}^{\II}$ is optimal for the given $\mv{\alpha}$, and $r_\II^\star(\mv{\alpha})< r_{\II\II}^\star(\mv{\alpha})$ otherwise. Therefore, the Pareto-optimal rate-pair $(R_1^\cc,R_2^\cc)$ along the rate-profile vector $\mv{\alpha}$ denoted by $(R_1^{\cc^\star}(\mv{\alpha}),R_2^{\cc^\star}(\mv{\alpha}))$ is given by $(\alpha_1,1-\alpha_1)\max(r_\II^\star(\mv{\alpha}),r_{\II\II}^\star(\mv{\alpha}))$. Moreover, we prove below that although only the SIC-achievable rate-pairs are considered in (P1), its optimal solutions $(R_1^{\cc^\star}(\mv{\alpha}),R_2^{\cc^\star}(\mv{\alpha}))$'s are able to fully characterize the capacity region $\mathcal{C}^\cc$.
		\begin{proposition}\label{prop_centralized}
			The capacity region of the IRS-aided two-user MAC in (\ref{channel_centralized}) under centralized IRS deployment is
			\begin{equation}\label{CR_cen}
				{\mathcal{C}}^\cc=\mathrm{Conv}\big((0,0){\bigcup}_{\mv{\alpha}:\alpha_1\in [0,1]} (R_1^{\cc^\star}(\mv{\alpha}),R_2^{\cc^\star}(\mv{\alpha})) \big).
			\end{equation}
		\end{proposition}
		\begin{IEEEproof}
			Please refer to Appendix \ref{proof_prop_centralized}.
		\end{IEEEproof}
		
		Proposition \ref{prop_centralized} indicates that the optimal solutions to (P1) with different $\mv{\alpha}$'s provide an alternative characterization of $\mathcal{C}^\cc$. However, (P1) is a non-convex optimization problem due to the uni-modular constraints on $\phi_m^\cc$'s in (\ref{P1c5}) as well as the complicated coupling among $p_1$, $p_2$, and $\mv{\Phi}^\cc$ in (\ref{P1c1})--(\ref{P1c2}). Therefore, finding the optimal solution to (P1) is generally a difficult task. In the following, we propose a high-quality suboptimal solution to (P1), based on which an inner bound of the capacity region can be obtained.
		
		\vspace{-4mm}
		\subsection{Capacity Region Inner Bound}\label{sec_cen_inner}
		\vspace{-1mm}
		In this subsection, we derive an inner bound of the capacity region $\mathcal{C}^\cc$ (or an achievable rate region) based on the rate-profile method presented above. Specifically, we propose an \emph{AO} based algorithm to find a high-quality suboptimal solution to the sum-rate maximization problem (P1) efficiently.
		
		First, we exploit the structure of (P1) to transform it into a more tractable form.
		\begin{proposition}\label{prop_P1}
			(P1) is equivalent to the following problem:
			\begin{align}
			\!\!\!\!\!	\mbox{(P2)}\!\!\! \underset{r,\mv{\Phi}^\cc: (\ref{P1c4}),(\ref{P1c5})}{\mathtt{max}}\!\!\! &r\\
				\mathtt{s.t.}\quad  &\log_2\left(1+\frac{P_{\pi_1}|\tilde{h}_{\pi_1}(\mv{\Phi}^\cc)|^2}{{2^{(1-\alpha_{\pi_1})r}\sigma^2}}\right)\geq \alpha_{\pi_1} r\label{P2c1}\\
				&\log_2\left(1+\frac{P_{\pi_2}|\tilde{h}_{\pi_2}(\mv{\Phi}^\cc)|^2}{\sigma^2}\right)\geq (1\!-\!\alpha_{\pi_1}) r.\!\!\!\!\label{P2c2}
			\end{align}
		\end{proposition}
		\begin{IEEEproof}
			Proposition \ref{prop_P1} can be proved by noting that the inequality in (\ref{P1c2}) can be replaced with equality without loss of optimality. We omit the details here for brevity.
		\end{IEEEproof}
		
		Note that for the case of $\alpha_{\pi_1}=1$, the optimal $\mv{\Phi}^\cc$ to (P2) can be readily derived as ${\phi}_{m}^\cc=e^{j(\arg\{\bar{h}_{\pi_1}\}-\arg\{{g}_{m}^\cc{h}_{{\pi_1}m}^\cc\})},\forall m$, similar to (\ref{D_phi}). Thus, we focus on solving (P2) with $\alpha_{\pi_1}\in [0,1)$ in the next. To further simplify (P2), we define an auxiliary variable $\beta\overset{\Delta}{=}2^{(1-\alpha_{\pi_1})r}$, which is an increasing function of $r$ for any $\alpha_{\pi_1}\in [0,1)$. (P2) is then equivalently rewritten as
		\begin{align}
			\mbox{(P3)} \underset{\beta,\mv{\Phi}^\cc: (\ref{P1c4}),(\ref{P1c5})}{\mathtt{max}}\quad &\beta\\
			\mathtt{s.t.}\qquad &|\tilde{h}_{\pi_1}^\cc(\mv{\Phi}^\cc)|^2\geq \frac{(\beta^{\frac{1}{1-\alpha_{\pi_1}}}-\beta)\sigma^2}{P_{\pi_1}}\label{P3c1}\\
			&|\tilde{h}_{\pi_2}^\cc(\mv{\Phi}^\cc)|^2\geq \frac{(\beta-1){\sigma^2}}{P_{\pi_2}}.\label{P3c2}
		\end{align}
		
		(P3) is still non-convex due to the uni-modular constraints on $\phi_m^\cc$'s as well as the quadratic terms on the left-hand sides (LHSs) of (\ref{P3c1}) and (\ref{P3c2}), for which the optimal solution is difficult to obtain. In the following, we adopt an \emph{AO} approach for finding a high-quality suboptimal solution to (P3). Specifically, note that each quadratic term $|\tilde{h}_k^\cc(\mv{\Phi}^\cc)|^2$ can be expressed as the following \emph{affine} form over each $\phi_m^\cc$ with the other reflection coefficients $\{\phi_i^\cc,i\neq m\}_{i=1}^M$ being fixed:
		\begin{equation}\label{affine}
			|\tilde{h}_k^\cc(\mv{\Phi}^\cc)|^2
			=2\mathfrak{Re}\{f_{2,km}\phi_m^\cc\}\! +\!f_{1,km},\quad k=1,2,
		\end{equation}
		where $f_{1,km}\overset{\Delta}{=}|\bar{h}_k+\sum_{i\neq m} g_i^\cc\phi_i^\cc h_{ki}^\cc|^2+|g_m^\cc h_{km}^\cc|^2$ and $f_{2,km}\overset{\Delta}{=}g_m^\cc h_{km}^\cc({\bar{h}_k}^*+\sum_{i\neq m} g_i^{\cc^*}\phi_i^{\cc^*}h_{ki}^{\cc^*})$, and the equality in (\ref{affine}) holds due to $|\phi_m^\cc|=1$. Hence, with given $\{\phi_i^\cc,i\neq m\}_{i=1}^M$, (P3) is reduced to the following problem:
		\begin{align}
		\!\!\!\!\!\!	\mbox{(P3-m)}\!\!\!\!\! \underset{\scriptstyle \beta \atop \scriptstyle \phi_m^\cc: |\phi_m^\cc|=1}{\mathtt{max}}\!\!\! &\beta\\[-2mm]
			\mathtt{s.t.}\ &2\mathfrak{Re}\{f_{2,\pi_1m}\phi_m^\cc\}\geq \frac{(\beta^{\frac{1}{1-\alpha_{\pi_1}}}-\beta)\sigma^2}{P_{\pi_1}}-f_{1,\pi_1m}\label{P3mc1}\\[-2mm]
			&2\mathfrak{Re}\{f_{2,\pi_2m}\phi_m^\cc\}\geq \frac{(\beta-1)\sigma^2}{P_{\pi_2}}-f_{1,\pi_2m}.\label{P3mc2}
		\end{align}
		
		Note that the only non-convexity in (P3-m) lies in the uni-modular constraint on $\phi_m^\cc$, thus motivating us to apply the \emph{convex relaxation} technique on this constraint. Specifically, we relax (P3-m) by replacing the constraint $|\phi_m^\cc|=1$ with a new convex constraint $|\phi_m^\cc|\leq 1$, and denote the relaxed problem as (P3-m-R). We then have the following proposition.
		\begin{proposition}\label{prop_relax}
			There exists an optimal solution of $\phi_m^\cc$ to (P3-m-R) that satisfies $|\phi_m^\cc|=1$.
		\end{proposition}
		\begin{IEEEproof}
			Please refer to Appendix \ref{proof_prop_relax}.
		\end{IEEEproof}
		
		Proposition \ref{prop_relax} indicates that the convex relaxation from (P3-m) to (P3-m-R) is \emph{tight}, and the optimal solution to (P3-m-R) that satisfies $|\phi_m^\cc|=1$ is also optimal for (P3-m). Thanks to the above transformations, (P3-m-R) is a convex optimization problem, which can be efficiently solved via the interior-point method with complexity $\mathcal{O}(1)$. If the obtained optimal solution does not satisfy $|\phi_m^\cc|=1$, another optimal solution with $|\phi_m^\cc|=1$ can be constructed via proper scaling and rotation according to Appendix \ref{proof_prop_relax}, with compleixty $\mathcal{O}(1)$. Therefore, by iteratively optimizing $(\beta,\phi_m^\cc)$ with all the other variables $\{\phi_i^\cc,i\neq m\}_{i=1}^M$ being fixed at each time via solving (P3-m), we can obtain a feasible solution to (P3) as well as (P1), which is in general suboptimal. An initial point for the above algorithm can be found by randomly generating $Q>1$ realizations of $\mv{\Phi}^\cc$ with the phase of each $\phi_m^\cc$ following uniform distribution in $[0,2\pi)$, and selecting the realization with the largest sum-rate under the given rate-profile vector. Note that since (P3-m) is solved \emph{optimally} in every iteration, the objective value of (P3), $\beta$, is non-decreasing over the iterations, which guarantees the \emph{monotonic convergence} of this algorithm since the sum-rate $r$ and hence $\beta$ is bounded above due to the finite transmit powers. For each $\mv{\alpha}$, let $\tilde{r}_\II(\mv{\alpha})$ and $\tilde{r}_{\II\II}(\mv{\alpha})$ denote the obtained solutions to (P1) with $\mv{\pi}=\mv{\pi}^{\II}$ and $\mv{\pi}=\mv{\pi}^{\II\II}$, respectively. Between their corresponding rate-pairs, we further select the one with larger sum-rate as
		\begin{equation} 
			(\tilde{R}_1^\cc(\mv{\alpha}),\tilde{R}_2^\cc(\mv{\alpha}))=(\alpha_1,1-\alpha_1)\max(\tilde{r}_\II(\mv{\alpha}),\tilde{r}_{\II\II}(\mv{\alpha})).
		\end{equation}
		By performing time sharing among the obtained $(\tilde{R}_1^\cc(\mv{\alpha}),\tilde{R}_2^\cc(\mv{\alpha}))$'s, an inner bound of the capacity region (or an achievable rate region) for the centralized IRS case is obtained as
		\begin{equation}\label{innerbound}
		\!\!\!	{\mathcal{C}}_\II^\cc=\mathrm{Conv}\Big((0,0) {\bigcup}_{\mv{\alpha}:\alpha_1\in [0,1]}(\tilde{R}_1^\cc(\mv{\alpha}),\tilde{R}_2^\cc(\mv{\alpha}))\Big)\subseteq \mathcal{C}^\cc.\!\!
		\end{equation}
		
		Note that the complexity for obtaining the above proposed solution to (P1) with both decoding orders can be shown to be $\mathcal{O}(2M(Q+I))$, where $I$ denotes the number of outer iterations (each requires solving (P3-m) for $M$ times from $m=1$ to $m=M$); moreover, the complexity for taking the convex hull of $L+1$ points in (\ref{innerbound}) is $\mathcal{O}(L\log L)$. Therefore, by approximating the $[0,1]$ range of the rate ratio $\alpha_1$ with $L$ uniformly sampled points, the overall complexity for obtaining ${\mathcal{C}}_\II^\cc$ is $\mathcal{O}(2M(Q+I)L+L\log L)$, which is \emph{polynomial} over $M$ and thus much lower than that of \hbox{the exhaustive search (i.e., $\mathcal{O}(L_0^M)$).}
		
		\vspace{-3mm}
		\subsection{Capacity Region Outer Bound}\label{sec_cen_outer}
		\vspace{-1mm}
		Next, we provide an outer bound of the capacity region $\mathcal{C}^\cc$. Specifically, it follows from (\ref{C1})--(\ref{C3}) that an outer bound of $\mathcal{C}^\cc$ can be constructed by finding an upper bound for each of $r_1^\cc(\mv{\Phi}^\cc)$, $r_2^\cc(\mv{\Phi}^\cc)$, and $r_{12}^\cc(\mv{\Phi}^\cc)$ separately, for which the details are given as follows.
		
		First, similar to (\ref{eq1}), it can be shown that for each user $k$, the effective channel gain $|\tilde{h}^{\cc}_k(\mv{\Phi}^\cc)|$ is upper-bounded by
		\begin{equation}\label{hmaxC}
			|\tilde{h}^{\cc}_k(\mv{\Phi}^\cc)|\leq |\bar{h}_k|+{\textstyle\sum}_{m=1}^M |g_m^\cc||h_{km}^\cc|\overset{\Delta}{=}\tilde{h}_{k,\ub}^{\cc},
		\end{equation}
		where the inequality holds with equality if and only if all the IRS reflection coefficients are designed to maximize user $k$'s effective channel gain, i.e.,
		\begin{equation}\label{phim}
			\phi_{m}^\cc=e^{j(\arg\{\bar{h}_k\}-\arg\{g_m^\cc h_{km}^\cc\})},\quad 
			m\in \mathcal{M}.
		\end{equation}
		Thus, based on (\ref{C1})--(\ref{C2}), each $r_k^\cc(\mv{\Phi}^\cc)$ is upper-bounded as
		\begin{equation}
			r_k^\cc(\mv{\Phi}^\cc)\leq \log_2(1+P_k\tilde{h}_{k,\ub}^{{\cc}^2}/\sigma^2)\overset{\Delta}{=}r_{k,\ub}^{\cc},\quad k=1,2.
		\end{equation}
		
		Next, we derive an upper bound for $r_{12}^\cc(\mv{\Phi}^\cc)$, which is a challenging task since $\mv{\Phi}^\cc$ can change both $\tilde{h}^{\cc}_1({\mv{\Phi}}^\cc)$ and $\tilde{h}^{\cc}_2(\mv{\Phi}^\cc)$ in $r_{12}^\cc(\mv{\Phi}^\cc)$. To achieve this goal, we formulate the following optimization problem:
		\begin{equation}
			\mbox{(P4)}\quad \underset{\mv{\Phi}^\cc:|\phi_m^\cc|=1,\forall m\in \mathcal{M}}{\mathtt{max}} P_1|\tilde{h}_1^\cc({\mv{\Phi}}^\cc)|^2+P_2|\tilde{h}_2^\cc({\mv{\Phi}}^\cc)|^2.
		\end{equation} 
		Let $s_0^\star$ denote the optimal value of (P4). Note that for any $s_0\geq s_0^\star$, $\log_2(1+s_0/\sigma^2)$ is an upper bound for $r_{12}^\cc(\mv{\Phi}^\cc)$. However, (P4) is a non-convex optimization problem due to the uni-modular constraints on $\phi_m^\cc$'s, thus $s_0^\star$ is generally difficult to obtain. In the following, we find an upper bound for $s_0^\star$ instead. First, we transform (P4) into a more tractable form. Define ${\mv{q}}_k^H\overset{\Delta}{=}{\mv{g}}^{\cc^T}\mathrm{diag}\{\mv{h}_k^\cc\}$, $\mv{v}\overset{\Delta}{=}P_1\bar{h}_1\mv{q}_1+P_2\bar{h}_2\mv{q}_2$, and $\mv{\phi}^\cc\overset{\Delta}{=}[\phi_1^\cc,...,\phi_M^\cc]^T$. Consequently, the objective function of (P4) can be rewritten as $P_1|\tilde{h}_1^\cc({\mv{\Phi}}^\cc)|^2+P_2|\tilde{h}_2^\cc({\mv{\Phi}}^\cc)|^2=P_1|\bar{h}_1|^2+P_2|\bar{h}_2|^2+{\mv{v}}^H\mv{\phi}^\cc
		+{\mv{\phi}}^{\cc^H}\mv{v}
		+\mv{\phi}^{\cc^H}(P_1\mv{q}_1\mv{q}_1^H+P_2\mv{q}_2\mv{q}_2^H)\mv{\phi}^\cc$, 
		which is a quadratic function of $\mv{\phi}^\cc$. Thus, we can apply the SDR technique for finding an upper bound for the optimal value of (P4). By introducing auxiliary variables ${\mv{w}}=[\mv{\phi}^{\cc^T},t]^T$ and $\mv{W}=\mv{ww}^H$, (P4) can be shown to be equivalent to the following problem with an additional constraint of $\mathrm{rank}(\mv{W})=1$:
		\begin{align}
			\mbox{(P4-SDR)}\quad \underset{\small{\mv{W}}}{\mathtt{max}}\quad &P_1|\bar{h}_1|^2+P_2|\bar{h}_2|^2+\mathrm{tr}\{\mv{WQ}\}\\
			\mathtt{s.t.}\quad &{\mv{W}}\succeq\mv{0}\\
			&[\small{\mv{W}}]_{m,m}=1, m=1,...,M+1,
		\end{align} 
		where $\mv{Q}\overset{\Delta}{=}[P_1\mv{q}_1\mv{q}_1^H+P_2\mv{q}_2\mv{q}_2^H,\mv{v};\mv{v}^H,0]$.
		(P4-SDR) is a semi-definite program (SDP) which can be efficiently solved via the interior-point method with complexity $\mathcal{O}(M^{6.5})$ \cite{SDR}. Denote $s^\star$ as the optimal value of (P4-SDR). Note that $s^\star\geq s_0^\star$ holds due to the relaxation of the rank-one constraint. Thus, we have $r_{12}^\cc(\mv{\Phi}^\cc)\leq \log_2(1+s^\star/\sigma^2)\overset{\Delta}{=}r_{12,\ub}^{\cc}$, which yields an outer bound of ${\mathcal{C}}^\cc$ given by
		\begin{align}
			{\mathcal{C}}^{\cc}_{\mathrm{O}}=\{(R_1^\cc,R_2^\cc):&R_1^\cc\leq r_{1,\ub}^{\cc},R_2^\cc\leq r_{2,\ub}^{\cc},\nonumber\\
			&R_1^\cc+R_2^\cc\leq r_{12,\ub}^{\cc}\}\supseteq{\mathcal{C}}^\cc.\label{OB2}
		\end{align}
		
		Besides the above bounds on the capacity region, we characterize the achievable rate regions with TDMA and FDMA for centralized IRS deployment as follows.
		
		\vspace{-3mm}
		\subsection{Achievable Rate Region with TDMA}\label{sec_MAC_C_TDMA}
		\vspace{-1mm}
		With TDMA, the centralized IRS should apply two different sets of reflection coefficients over the two time slots, each tailored for one of the two users without loss of generality. Let $\mv{\Phi}^\cc[k]$ denote the reflection coefficients at the time slot allocated to the $k$th user. For any given $\{\mv{\Phi}^\cc[k]\}$, the TDMA achievable rate region is given by $
		\mathcal{R}^{\cc}_{\ttt}(\{\mv{\Phi}^{\cc}[k]\})={\bigcup}_{\rho_{\ttt}\in [0,1]}\{(R_1^\cc,R_2^\cc):
		R_1^\cc\leq \rho_{\ttt}\log_2(1+P_1|\tilde{h}_1^\cc(\mv{\Phi}^{\cc}[1])|^2/\sigma^2),
		R_2^\cc\leq (1-\rho_{\ttt})\log_2(1+P_2|\tilde{h}_2^\cc(\mv{\Phi}^{\cc}[2])|^2/\sigma^2)\}$. Note that the channel gain of the $k$th user at its assigned time slot,  $|\tilde{h}_k^\cc(\mv{\Phi}^{\cc}[k])|$, can be maximized as $\tilde{h}_{k,\ub}^\cc$ by setting $\mv{\Phi}^{\cc}[k]$ as $\mv{\Phi}^\cc$ given in (\ref{phim}). Hence, similar to the distributed IRS case, the TDMA achievable rate region with centralized IRS is given by
		\begin{align}
			\mathcal{R}^{\cc}_{\ttt}=&{\bigcup}_{\rho_{\ttt}\in [0,1]}\Bigg\{(R_1^\cc,R_2^\cc):
			R_1^\cc\leq \rho_{\ttt}\log_2\left(1+\frac{P_1\tilde{h}_{1,\ub}^{\cc^2}}{\sigma^2}\right),\nonumber\\
			&R_2^\cc\leq (1-\rho_{\ttt})\log_2\left(1+\frac{P_2\tilde{h}_{2,\ub}^{\cc^2}}{\sigma^2}\right)\Bigg\}.\label{T_C}
		\end{align}
	
		\vspace{-3mm}
		\subsection{Achievable Rate Region with FDMA}
		\vspace{-1mm}
		With FDMA, the achievable rate region for any given $\mv{\Phi}^{\cc}$ is given by $\mathcal{R}^{\cc}_{\ff}(\mv{\Phi}^{\cc})={\bigcup}_{\rho_{\ff}\in [0,1]}\{(R_1^\cc,R_2^\cc):
		R_1^\cc\leq \rho_{\ff}\log_2(1\!+\!P_1|\tilde{h}_1^\cc(\mv{\Phi}^\cc)|^2/(\rho_{\ff}\sigma^2)),R_2^\cc\leq (1-\rho_{\ff})\log_2(1\!+\!P_2|\tilde{h}_2^\cc(\mv{\Phi}^\cc)|^2/((1\!-\!\rho_{\ff})\sigma^2))\}$. 
		After time sharing among different $\mv{\Phi}^{\cc}$'s, the overall achievable rate region is given by $\mathcal{R}^{\cc}_{\ff}=\mathrm{Conv}\Big({\bigcup}_{{\mv{\Phi}}^\cc\in \mathcal{F}^\cc} {\mathcal{R}}_{\ff}^\cc({\mv{\Phi}}^\cc)\Big)$. Similar to the characterization of $\mathcal{C}^{\cc}$ in Section \ref{sec_cen_cha}, $\mathcal{R}^{\cc}_{\ff}$ can be characterized via the rate-profile based method by solving the following problem for $\alpha_1\in [0,1]$:
		\begin{align}
		\mbox{(P5)} \!\!\!\!\!\!
		\underset{\scriptstyle r,\rho_\ff \atop \scriptstyle \mv{\Phi}^\cc: (\ref{P1c4}),(\ref{P1c5})}{\mathtt{max}}\!\!\! &r\\
			\mathtt{s.t.}\quad  &\rho_\ff\log_2\left(1+ \frac{P_1|\tilde{h}_1^\cc(\mv{\Phi}^\cc)|^2}{\rho_\ff\sigma^2}\right)\geq \alpha_1 r\label{P4c1}\\
			&(1-\rho_\ff)\log_2\left(1+\frac{P_2|\tilde{h}_2^\cc(\mv{\Phi}^\cc)|^2}{(1-\rho_\ff)\sigma^2}\right)\geq (1-\alpha_1) r\label{P4c2}\\
			& 0\leq \rho_\ff\leq 1.\label{P4c3}
		\end{align}
				
		Note that (P5) is a non-convex optimization problem due to the complicated coupling between $\rho_\ff$ and $\mv{\Phi}^\cc$. Nevertheless, a suboptimal solution to (P5) can be found via AO in a similar manner as that for (P1). For brevity, the details are given in Appendix \ref{prop_P5}. Based on this, an inner bound of $\mathcal{R}_\ff^\cc$, denoted as $\mathcal{R}_{\ff,\II}^\cc$, can be similarly obtained as $\mathcal{C}_\II^\cc$. On the other hand, we have $\mathcal{R}_\ff^\cc(\mv{\Phi}^\cc)\subseteq \mathcal{C}^\cc(\mv{\Phi}^\cc)$ for any $\mv{\Phi}^\cc$ \cite{Elements}, and consequently $\mathcal{R}_\ff^\cc\subseteq \mathcal{C}^\cc$. Therefore, the outer bound for the capacity region $\mathcal{C}^\cc$, $\mathcal{C}_{\OO}^\cc$, is also an outer bound of $\mathcal{R}_\ff^\cc$, i.e., $\mathcal{R}_\ff^\cc\subseteq \mathcal{R}_{\ff,\OO}^\cc\overset{\Delta}{=}\mathcal{C}_{\OO}^\cc$.
		
		Finally, we show that $\mathcal{R}_\ttt^\cc\subseteq\mathcal{R}_\ff^\cc$. Let $\mv{\Phi}^\cc_{(k)}$ denote the IRS reflection coefficients that achieve the maximum effective channel gain for the $k$th user, $\tilde{h}_{k,\ub}^{\cc}$, which are given in (\ref{phim}). It can be easily shown that time sharing of the FDMA achievable rate regions with $\mv{\Phi}^\cc_{(1)}$ and $\mv{\Phi}^\cc_{(2)}$ suffices to contain the TDMA achievable rate region due to the curved Pareto boundary of $\mathcal{R}_\ff^\cc(\mv{\Phi}^\cc)$. Therefore, we have $\mathcal{R}_{\ttt}^\cc\subseteq\mathrm{Conv}\Big({\mathcal{R}}_{\ff}^\cc({\mv{\Phi}}_{(1)}^\cc)\cup{\mathcal{R}}_{\ff}^\cc({\mv{\Phi}}_{(2)}^\cc)\Big)\subseteq \mathcal{R}^{\cc}_{\ff}$, namely, the FDMA achievable rate region contains the TDMA achievable rate region. It is worth noting that although with any given $\mv{\Phi}^\cc$, the FDMA achievable rate region may not contain that of TDMA, time sharing over different $\mv{\Phi}^\cc$'s enables FDMA to outperform TDMA in terms of achievable rate region. 
		
		\vspace{-4mm}
		\section{Performance Comparison: Distributed vs. Centralized IRS Deployment}\label{sec_comparison}
		\vspace{-1mm}
		In this section, we compare the capacity regions and TDMA/FDMA achievable rate regions under the two IRS deployment strategies. For simplicity, we assume that the direct user-AP channels are negligible as compared to the IRS-reflected channels and thus $\bar{h}_1=\bar{h}_2=0$, which is practically valid for IRSs with large $M$ (and thus $M_1$ and $M_2$).\footnote{The general case with non-zero $\bar{h}_1$ and $\bar{h}_2$ is more difficult to analyze, which is thus considered for the numerical example in Section \ref{sec_num}.} Moreover, for fair comparison, we consider the following \emph{twin} channels (defined in Assumption 1 below) between the two deployment cases, where the two distributed user-IRS channels constitute the centralized IRS-AP channel, and each user-IRS channel in the centralized case contains the corresponding IRS-AP channel in the distributed case.
		\begin{assumption}[Twin Channels]
			For the channel coefficients illustrated in Fig. \ref{fig_system}, we assume $\mv{g}^\cc=[\mv{h}_1^{\dd^T},\mv{h}_2^{\dd^T}]^T$, $h_{1m}^\cc=g_{1m}^\dd,\ \forall m\in \mathcal{M}_1$, and $h_{2(m+M_1)}^\cc=g_{2m}^\dd,\ \forall m\in \mathcal{M}_2$.
		\end{assumption}
		
		The above twin channels hold in practice if the user-IRS channels in the distributed case have the same statistical distribution and link distance as the IRS-AP channel in the centralized case, and the IRS-AP channels in the distributed case have the same statistical distribution and link distance as the corresponding user-IRS channels in the centralized case (see Fig. \ref{fig_system}).\footnote{For scenarios where the two deployment strategies may lead to different channel statistical distributions, their capacity comparison is more complicated and difficult to be performed analytically, which will be further discussed in Section \ref{sec_con} to motivate future work.}
		
		\vspace{-3mm}
		\subsection{Capacity Region Comparison}
				\vspace{-1mm}
		First, we have the following proposition for the capacity region comparison.
		\begin{proposition}\label{prop_compare}
			Under $\bar{h}_1=\bar{h}_2=0$ and Assumption 1, the capacity region of the centralized IRS deployment contains that of the distributed IRS deployment, i.e., $\mathcal{C}^\dd \subseteq \mathcal{C}^\cc$.
		\end{proposition}
		\begin{IEEEproof}
			We construct $\tilde{\mv{\Phi}}^\cc$ for the centralized IRS such that the reflection coefficients of its two sub-surfaces, $\{\tilde{\phi}_m^\cc\}_{m=1}^{M_1}$ and $\{\tilde{\phi}_m^\cc\}_{m=M_1+1}^M$,  correspond to the capacity-achieving reflection coefficients at IRS 1 and 2 for the distributed deployment shown in (\ref{D_phi}), respectively, but each being rotated by a common phase $\theta_1\in [0,2\pi)$ or $\theta_2\in [0,2\pi)$, i.e.,
			\begin{equation}\label{phase1}
				\tilde{\phi}_m^\cc=
				\begin{cases}
					e^{j(\arg\{\bar{h}_1\}-\arg\{g_{1m}^\dd h_{1m}^\dd\}+\theta_1)},\!\!\!&m\in \mathcal{M}_1\\
					e^{j(\arg\{\bar{h}_2\}-\arg\{g_{2(m\!-\!M_1)}^\dd h_{2(m\!-\!M_1)}^\dd\}+\theta_2)},\!\!\!\!&m\!\in\! \mathcal{M}\backslash\mathcal{M}_1.\!\!\!\!\!\!
				\end{cases}
			\end{equation}
			Then, we have $|\tilde{h}_k^\cc(\tilde{\mv{\Phi}}^\cc)|=|\tilde{h}_{k,\ub}^\dd+\tilde{f}_ke^{j(\theta_2-\theta_1)}|$, with $\tilde{f}_1=\sum_{m=M_1+1}^M g_m^\cc h_{1m}^\cc e^{-j\arg\{g_{2(m-M_1)}^\dd h_{2(m-M_1)}^\dd\}}$ and $\tilde{f}_2=\sum_{m=1}^{M_1} g_m^{\cc^*} h_{2m}^{\cc^*} e^{j\arg\{g_{1m}^\dd h_{1m}^\dd\}}$ (recall $\tilde{h}_{k,\ub}^\dd$'s are the capacity-achieving effective channel gains for the distributed case). Then, we prove that we can always design $\theta_1$ and $\theta_2$ such that $|\tilde{h}_k^\cc(\tilde{\mv{\Phi}}^\cc)|\geq\tilde{h}_{k,\ub}^\dd$ holds for any $k\in \{1,2\}$. To this end, we present Lemma \ref{lemma_phase} as below.
			\begin{lemma}\label{lemma_phase}
				For any complex numbers $\{a_k,b_k\}_{k=1}^2$, denote $c_k=\arg\{b_k\}-\arg\{a_k\}\in [0,2\pi)$, $k=1,2$. Then,
				$|a_k+b_ke^{j\theta}|\geq |a_k|,\ k=1,2$ holds with $\theta=\frac{\pi}{2}-\min(c_1,c_2)$ if $|c_1-c_2|\geq \pi$, and $\theta=\frac{\pi}{2}-\max(c_1,c_2)$ otherwise.
			\end{lemma}
			\begin{IEEEproof}
				Please refer to Appendix \ref{proof_lemma_phase}.
			\end{IEEEproof}
			
			By substituting $a_k$ and $b_k$ with $\tilde{h}_{k,\ub}^\dd$ and $\tilde{f}_k$, respectively, and setting $\theta_2-\theta_1$ as $\theta$ in Lemma \ref{lemma_phase}, we have $|\tilde{h}_k^\cc(\tilde{\mv{\Phi}}^\cc)|\geq\tilde{h}_{k,\ub}^\dd$ for any $k\in \{1,2\}$, and consequently $\mathcal{C}^\dd\subseteq \mathcal{C}^\cc(\tilde{\mv{\Phi}}^\cc)\subseteq \mathcal{C}^\cc$. This thus completes the proof of Proposition \ref{prop_compare}.
		\end{IEEEproof}
		
		Proposition \ref{prop_compare} indicates that by judiciously designing the IRS reflection, the larger passive beamforming gain at the centralized IRS can benefit the two users at the same time, thus yielding a larger capacity region than the case with two distributed IRSs each serving one user only.

		\newcommand{\tabincell}[2]{\begin{tabular}{@{}#1@{}}#2\end{tabular}}  
\begin{table*}[t]
	\centering
	\caption{Summary of Main Results for IRS-Aided Two-User MAC}\label{table_MAC}
	\begin{tabular}{|c|c|c|c|}
		\hline 
		&	Distributed	& Centralized &  Comparison (Twin Channels)\\ 
		\hline
		Capacity Region & $\mathcal{C}^\dd$ (closed-form) & \tabincell{c}{$\mathcal{C}_{\II}^\cc$ (rate-profile); $\mathcal{C}_{\OO}^\cc$ (SDR)\\ $\mathcal{C}_{\II}^\cc\subseteq\mathcal{C}^\cc\subseteq \mathcal{C}_{\OO}^\cc$} & \tabincell{c}{
			$\mathcal{C}^\dd\subseteq \mathcal{C}^\cc$ (with $\bar{h}_k=0,\ \forall k$)} \\
		\hline
		\tabincell{c}{Achievable Rate\\ Region with TDMA} & $\mathcal{R}^\dd_\ttt$ (closed-form) & $\mathcal{R}^\cc_\ttt$ (closed-form) & $\mathcal{R}^\dd_\ttt\subseteq \mathcal{R}^\cc_\ttt$ \\
		\hline 
		\tabincell{c}{Achievable Rate\\ Region with FDMA} & \tabincell{c}{
			$\mathcal{R}^\dd_\ff$ (closed-form)\\
			$\mathcal{R}^\dd_\ttt\subseteq\mathcal{R}^\dd_\ff\subseteq \mathcal{C}^\dd$ } & \tabincell{c}{$\mathcal{R}^\cc_{\ff,\II}$ (rate-profile); $\mathcal{R}^\cc_{\ff,\OO}$ \\
			$\mathcal{R}^\cc_{\ff,\II}\subseteq \mathcal{R}^\cc_{\ff}\subseteq \mathcal{R}^\cc_{\ff,\OO}$\\
			$\mathcal{R}^\cc_\ttt\subseteq\mathcal{R}^\cc_\ff\subseteq \mathcal{C}^\cc$ } & $\mathcal{R}_\ff^\dd\subseteq \mathcal{R}_\ff^\cc$ (with $\bar{h}_k=0,\ \forall k$)\\
		\hline
	\end{tabular}
	
	\vspace{-3mm}
\end{table*}
		
		\vspace{-3mm}
		\subsection{Achievable Rate Region Comparison with TDMA and FDMA}
				\vspace{-1mm}
		For TDMA, note from (\ref{T_D}) and (\ref{T_C}) that the expression for $\mathcal{R}_\ttt^\cc$ is the same as $\mathcal{R}_\ttt^\dd$ by replacing $\tilde{h}_{k,\ub}^\dd$ with $\tilde{h}_{k,\ub}^\cc$. Under the twin channel condition specified in Assumption 1, it can be shown from (\ref{eq1}) and (\ref{hmaxC}) that the maximum effective channel gain for each user under centralized deployment (i.e., $\tilde{h}_{k,\ub}^\cc$) is larger than that under distributed deployment (i.e., $\tilde{h}_{k,\ub}^\dd$) due to the larger-size IRS available for passive beamforming, thus we have $\mathcal{R}_\ttt^\dd\subseteq \mathcal{R}_\ttt^\cc$, i.e., the TDMA achievable rate region under centralized deployment contains that under distributed deployment.
			
		For FDMA, recall from the proof of Proposition \ref{prop_compare} that we can always construct $\tilde{\mv{\Phi}}^\cc$ such that $|\tilde{h}_k^\cc(\tilde{\mv{\Phi}}^\cc)|\geq \tilde{h}_{k,\ub}^\dd$ holds for both $k=1$ and $k=2$ under the twin channel condition in Assumption 1 and negligible direct channels $\bar{h}_1=\bar{h}_2=0$. Therefore, it follows that $\mathcal{R}_\ff^\dd\subseteq \mathcal{R}_\ff^\cc(\tilde{\mv{\Phi}}^\cc)\subseteq \mathcal{R}_\ff^\cc$ holds, i.e., the FDMA achievable rate region under centralized deployment contains that under distributed deployment under Assumption 1 and $\bar{h}_1=\bar{h}_2=0$.
		
		The above results indicate that the superior rate performance of centralized IRS deployment over distributed IRS deployment still holds for practical OMA schemes (i.e., TDMA/FDMA) under the assumed channel conditions. For ease of reference, we summarize in Table \ref{table_MAC} our main results on the capacity/rate region characterization and comparison for the two-user MAC.
		
		\vspace{-3mm}
		\section{Extension to IRS-Aided Two-User BC}\label{sec_BC}
		In this section, we extend our capacity/rate region characterization of the uplink IRS-aided MAC to the downlink IRS-aided BC, by leveraging the celebrated \emph{uplink-downlink duality} (or MAC-BC duality) framework \cite{Elements}. For ease of exposition, we consider a dual channel setup where all the downlink channels equal to their uplink counterparts, thus the effective channels from the AP to user $k$ under distributed and centralized deployment are the corresponding effective user-AP channels in the MAC case, i.e., $\tilde{h}_k^\dd(\mv{\Phi}_k^\dd)$ and $\tilde{h}_k^\cc(\mv{\Phi}^\cc)$ given in (\ref{channel_distributed}) and (\ref{channel_centralized}), respectively.
		
		\vspace{-4mm}
		\subsection{Distributed IRS Deployment}\label{sec_BC_dis}
		Under distributed IRS deployment, the received signal at the $k$th user is given by
		\begin{equation}\label{signal_dis_BC}
			y_k=\tilde{h}_k^\dd(\mv{\Phi}_k^\dd)x+z_k,\quad k=1,2,
		\end{equation}
		where $x=\sqrt{p_1}s_1+\sqrt{p_2}s_2$ denotes the transmitted signal at the AP, with $p_k$ and $s_k$ denoting the transmit power and information symbol for user $k$, respectively; $z_k\sim \mathcal{CN}(0,\sigma^2)$ denotes the receiver noise at user $k$. We consider a transmit power constraint $P$ at the AP, thus we have $p_1+p_2\leq P$. For any given IRS reflection coefficients $\{\mv{\Phi}^\dd_k\}$, it follows from the uplink-downlink duality that the capacity region of the two-user BC equals to the union set of its dual MAC capacity regions with transmit power constraint pairs $(P_1,P_2)$'s that satisfy $P_1+P_2=P$ \cite{Elements}, which is given by
		\begin{equation}
			\mathcal{C}^\dd_{\BC}(\{\mv{\Phi}^\dd_k\})={\bigcup}_{(P_1,P_2):P_1+P_2= P} \mathcal{C}^\dd(\{\mv{\Phi}^\dd_k\}).
		\end{equation}
		By considering time sharing among different $\{\mv{\Phi}^\dd_k\}$'s, the overall capacity region is given by
		\begin{align}
			&\mathcal{C}^\dd_{\BC}=\mathrm{Conv}\Big({\bigcup}_{\{{\mv{\Phi}}^\dd_k\}\in \mathcal{F}^\dd} \mathcal{C}^\dd_{\BC}(\{\mv{\Phi}^\dd_k\})\Big)\nonumber\\
			=&\mathrm{Conv}\Big({\bigcup}_{\{{\mv{\Phi}}^\dd_k\}\in \mathcal{F}^\dd}{\bigcup}_{(P_1,P_2):P_1+P_2= P} \mathcal{C}^\dd(\{\mv{\Phi}^\dd_k\})\Big)\nonumber\\
			=&\mathrm{Conv}\Big({\bigcup}_{(P_1,P_2):P_1+P_2= P}\mathrm{Conv}\Big({\bigcup}_{\{{\mv{\Phi}}^\dd_k\}\in \mathcal{F}^\dd} \mathcal{C}^\dd(\{\mv{\Phi}^\dd_k\})\Big)\Big)\nonumber\\
			=&\mathrm{Conv}\Big({\bigcup}_{(P_1,P_2):P_1+P_2= P}\mathcal{C}^\dd\Big).\label{C_dis_BC}
		\end{align}
		Note that the capacity-achieving optimal input distribution for any given $\{\mv{\Phi}^\dd_k\}$ is the CSCG distribution with $s_k\sim \mathcal{CN}(0,1),\ \forall k$, which is thus also optimal for the IRS-aided BC under distributed deployment according to (\ref{C_dis_BC}) and will be assumed in the sequel. For the purpose of exposition, we define $\alpha_\PP\overset{\Delta}{=}\frac{P_1}{P}\in [0,1]$. Recall from Section \ref{sec_dis} that the capacity region for the dual MAC, $\mathcal{C}^\dd$, is derived in closed-form in Theorem \ref{theorem_dis}. Hence, $\mathcal{C}^\dd_{\BC}$ can be characterized based on (\ref{C_dis_BC}) by obtaining $\mathcal{C}^\dd$ for every $(P_1,P_2)$ with $P_1+P_2=P$ via one-dimensional search over $\alpha_\PP$, and then taking the convex hull of their union set. By approximating the $[0,1]$ range of the power ratio $\alpha_\PP$ with $L_\PP$ uniformly sampled points, the overall complexity for obtaining $\mathcal{C}^\dd_{\BC}$ is $\mathcal{O}(L_\PP\log(L_\PP))$, which is dominated by the convex hull operation.
		
		Similarly, the achievable rate regions with TDMA and FDMA can be characterized based on their MAC counterparts $\mathcal{R}_{\ttt}^\dd$ and $\mathcal{R}_{\ff}^\dd$ (which are available in closed-form) as $\mathcal{R}^\dd_{\BC,\ttt}=\mathrm{Conv}\Big({\bigcup}_{(P_1,P_2):P_1+P_2= P}\mathcal{R}_{\ttt}^\dd\Big)$ and $\mathcal{R}^\dd_{\BC,\ff}=\mathrm{Conv}\Big({\bigcup}_{(P_1,P_2):P_1+P_2= P}\mathcal{R}_{\ff}^\dd\Big)$, respectively. Note that since $\mathcal{R}^\dd_\ttt\subseteq\mathcal{R}^\dd_\ff\subseteq \mathcal{C}^\dd$ holds for MAC, we have $\mathcal{R}^\dd_{\BC,\ttt}\subseteq\mathcal{R}^\dd_{\BC,\ff}\subseteq \mathcal{C}_{\BC}^\dd$ for the dual BC.
		
		\vspace{-4mm}
		\subsection{Centralized IRS Deployment}\label{sec_BC_cen}
		Under centralized IRS deployment, the received signal at each $k$th user is given by 
		\begin{equation}\label{signal_cen_BC}
			y_k=\tilde{h}_k^\cc(\mv{\Phi}^\cc)x+z_k,\quad k=1,2.
		\end{equation}
		By leveraging the uplink-downlink duality, the capacity region of the two-user BC with any given IRS reflection coefficients $\mv{\Phi}^\cc$ is given by $\mathcal{C}_{\BC}^\cc(\mv{\Phi}^\cc)={\bigcup}_{(P_1,P_2):P_1+P_2= P} \mathcal{C}^\cc(\mv{\Phi}^\cc)$. Similar to the distributed IRS case, the BC capacity region can be obtained by considering time sharing among different $\mv{\Phi}^\cc$'s and thus expressed in terms of the dual MAC capacity region $\mathcal{C}^\cc$ as follows:
		\begin{align}\label{C_central_BC}
			\mathcal{C}^\cc_{\BC}=&\mathrm{Conv}\Big({\bigcup}_{{\mv{\Phi}}^\cc\in \mathcal{F}^\cc} \mathcal{C}^\cc_{\BC}(\mv{\Phi}^\cc)\Big)\nonumber\\
			=&\mathrm{Conv}\Big({\bigcup}_{(P_1,P_2):P_1+P_2= P}\mathcal{C}^\cc\Big).
		\end{align}
		Notice from (\ref{C_central_BC}) that the optimal input distribution for IRS-aided BC under the centralized deployment is still $s_k\sim \mathcal{CN}(0,1),\ \forall k$ due to its optimality for any given ${\mv{\Phi}}^\cc$, which is thus assumed in the sequel.
		Since no closed-form expression is available for $\mathcal{C}^\cc$, we find inner and outer bounds for $\mathcal{C}^\cc_{\BC}$ based on those for $\mathcal{C}^\cc$ in the following.
		\subsubsection{Capacity Region Inner Bound}
		Recall from Proposition \ref{prop_centralized} that $\mathcal{C}^\cc$ can be characterized via the rate-profile method by solving a series of sum-rate maximization problems in (P1) with different rate-profile vectors $\mv{\alpha}$. Motivated by this and based on (\ref{C_central_BC}), we take a similar approach to characterize ${\mathcal{C}}^\cc_{\BC}$. Specifically, we formulate the following sum-rate maximization problem with given rate-profile vector $\mv{\alpha}$ and decoding order $\mv{\pi}$ by replacing the individual transmit power constraints in (P1) for the MAC case with a sum transmit power constraint:
		\begin{align}
			\mbox{(P6)}\quad \underset{r,p_1,p_2,\mv{\Phi}^\cc: (\ref{P1c1}), (\ref{P1c2}), (\ref{P1c4}), (\ref{P1c5})}{\mathtt{max}}\ &r\\
			\mathtt{s.t.}\qquad\qquad & p_1+p_2\leq P. \label{P6c3}
		\end{align}
		
		For each $\mv{\alpha}$, let ${r}^\star_{\BC,\II}(\mv{\alpha})$ and ${r}^\star_{\BC,\II\II}(\mv{\alpha})$ denote the optimal solutions to (P6) with $\mv{\pi}=\mv{\pi}^\II$ and $\mv{\pi}=\mv{\pi}^{\II\II}$, respectively. Similar to the MAC case elaborated in Section \ref{sec_cen}-A, the directly achievable Pareto-optimal rate-pair (without the need of time sharing/rate splitting) for the two-user BC along the rate-profile vector $\mv{\alpha}=[\alpha_1,1-\alpha_1]^T$ is given by
		\begin{align} \label{BC_pair}
			&(R_{\BC,1}^{\cc^\star}(\mv{\alpha}),R_{\BC,2}^{\cc^\star}(\mv{\alpha}))\nonumber\\
			=&
			(\alpha_1,1-\alpha_1)\max({r}^\star_{\BC,\II}(\mv{\alpha}),{r}^\star_{\BC,\II\II}(\mv{\alpha})).
		\end{align}
		The following proposition then follows directly from the above and (\ref{C_central_BC}).
		\begin{proposition}\label{prop_centralized_BC}
			The capacity region of the IRS-aided two-user BC with centralized IRS deployment is
			\begin{equation}
			\!\!\!\!	{\mathcal{C}}^\cc_{\BC}=\mathrm{Conv}\big((0,0){\bigcup}_{\mv{\alpha}:\alpha_1\in [0,1]} (R_{\BC,1}^{\cc^\star}(\mv{\alpha}),R_{\BC,2}^{\cc^\star}(\mv{\alpha})) \big).\!\!
			\end{equation}
		\end{proposition}
		
		Next, we proceed to solve (P6). Similar to (P1), (P6) is a non-convex optimization problem whose optimal solution is difficult to obtain. Thus, we adopt the AO technique to find a suboptimal solution for it. Specifically, we define an auxiliary variable $\beta\overset{\Delta}{=}2^{(1-\alpha_{\pi_1})r}$. With any given $\mv{\Phi}^\cc$, (P6) is equivalent to the following optimization problem over $(\beta,p_1,p_2)$:
		\begin{align}
			\mbox{(P6-P)} \underset{\beta,p_1,p_2:p_1+p_2\leq P}{\mathtt{max}}\ &\beta\\
			\mathtt{s.t.}\qquad &\frac{p_{\pi_1}|\tilde{h}_{\pi_1}^\cc(\mv{\Phi}^\cc)|^2}{\sigma^2}\geq \beta^{\frac{1}{1-\alpha_{\pi_1}}}-\beta \label{P6Pc1}\\
			&\frac{p_{\pi_2}|\tilde{h}_{\pi_2}^\cc(\mv{\Phi}^\cc)|^2}{\sigma^2}\geq \beta-1\label{P6Pc2}.
		\end{align}
		
		(P6-P) is a convex optimization problem that can be solved efficiently with complexity $\mathcal{O}(1)$. On the other hand, with any given $p_1,p_2$ and $\{\phi_i^\cc,i\neq m\}_{i=1}^M$, (P6) is equivalent to (P3-m) by replacing each $P_{\pi_k}$ with $p_{\pi_k}$, $k=1,2$, which is denoted as (P6-m). The optimal solution to (P6-m) can be obtained similarly as that of (P3-m) with complexity $\mathcal{O}(1)$. Therefore, similar to the proposed algorithm for (P1), by iteratively optimizing $(\beta,p_1,p_2)$ or $(\beta,\phi_m^\cc)$ for one element $m\in \mathcal{M}$ with all the other optimization variables being fixed at each time, a feasible solution to (P6) can be obtained, which is generally suboptimal. Note that \emph{monotonic convergence} is guaranteed for the proposed algorithm since the \emph{optimal} solution is found for every sub-problem, and the maximum sum-rate is bounded above. For each $\mv{\alpha}$, let $\tilde{r}_{\BC,\II}(\mv{\alpha})$ and $\tilde{r}_{\BC,\II\II}(\mv{\alpha})$ denote the obtained solutions to (P6) with $\mv{\pi}=\mv{\pi}^{\II}$ and $\mv{\pi}=\mv{\pi}^{\II\II}$, respectively. We further select the one with larger sum-rate between their corresponding rate-pairs as $(\tilde{R}_{\BC,1}^\cc(\mv{\alpha}),\tilde{R}_{\BC,2}^\cc(\mv{\alpha}))$ in a similar manner as the selection of $(R_{\BC,1}^{\cc^\star}(\mv{\alpha}),R_{\BC,2}^{\cc^\star}(\mv{\alpha}))$ in (\ref{BC_pair}) by replacing ${r}^\star_{\BC,\II}(\mv{\alpha})$ and ${r}^\star_{\BC,\II\II}(\mv{\alpha})$ with $\tilde{r}_{\BC,\II}(\mv{\alpha})$ and $\tilde{r}_{\BC,\II\II}(\mv{\alpha})$, respectively. By performing time sharing among the obtained $(R_{\BC,1}^{\cc^\star}(\mv{\alpha}),R_{\BC,2}^{\cc^\star}(\mv{\alpha}))$'s, an inner bound of the two-user BC capacity region (or an achievable rate region) is obtained as
		\begin{align}
			{\mathcal{C}}_{\BC,\II}^\cc=&\mathrm{Conv}\Big((0,0) {\bigcup}_{\mv{\alpha}:\alpha_1\in [0,1]}(\tilde{R}_{\BC,1}^\cc(\mv{\alpha}),\tilde{R}_{\BC,2}^\cc(\mv{\alpha}))\Big)\nonumber\\
			\subseteq& \mathcal{C}_\BC^\cc.
		\end{align}  
		The complexity for obtaining ${\mathcal{C}}_{\BC,\II}^\cc$ can be shown to be $\mathcal{O}(2M(Q_{\BC}+I_{\BC})L+L\log L)$, with $Q_{\BC}$ denoting the number of random realizations of $\mv{\Phi}^\cc$ in the initialization, $I_{\BC}$ denoting the number of outer iterations (each requiring solving (P6-P) once and (P6-m) for $M$ times), and $L$ denoting the number of points for approximating the rate ratio range $[0,1]$.
		
		\subsubsection{Capacity Region Outer Bound}
		On the other hand, recall from Section \ref{sec_cen_outer} that an outer bound for $\mathcal{C}^\cc$, $\mathcal{C}_\OO^\cc$, can be obtained with complexity $\mathcal{O}(M^{6.5})$. Based on this and (\ref{C_central_BC}), we have $\mathcal{C}^\cc_{\BC}\subseteq \mathrm{Conv}\Big({\bigcup}_{(P_1,P_2):P_1+P_2= P}\mathcal{C}_\OO^\cc\Big)\overset{\Delta}{=}\mathcal{C}^\cc_{\BC,\OO}$, thus an outer bound for $\mathcal{C}^\cc_{\BC}$, $\mathcal{C}^\cc_{\BC,\OO}$, can be characterized by obtaining $\mathcal{C}_\OO^\cc$ for all $(P_1,P_2)$'s that satisfy $P_1+P_2=P$ and taking the convex hull of their union set. The required complexity is $\mathcal{O}(L_\PP M^{6.5}+L_\PP\log L_\PP)$ by approximating the $[0,1]$ range of $\alpha_{\PP}=\frac{P_1}{P}$ with $L_\PP$ uniformly sampled points.
		\subsubsection{Achievable Rate Regions with TDMA and FDMA}
		Similar to the distributed IRS case, the achievable rate region with TDMA for the case of centralized IRS can be characterized as $\mathcal{R}^\cc_{\BC,\ttt}=\mathrm{Conv}\Big({\bigcup}_{(P_1,P_2):P_1+P_2= P}\mathcal{R}_{\ttt}^\cc\Big)$, by leveraging the closed-form expression of $\mathcal{R}_{\ttt}^\cc$ derived in Section \ref{sec_MAC_C_TDMA}. For FDMA, we characterize inner and outer bounds of the achievable rate region (denoted as $\mathcal{R}_{\BC,\ff}^\cc$) similarly as those for the MAC case. Specifically, an inner bound of $\mathcal{R}_{\BC,\ff}^\cc$ (denoted by $\mathcal{R}_{\BC,\ff,\II}^\cc$) can be found via the rate-profile method by solving a series of sum-rate maximization problems with different rate-profile vectors, where each problem is an extended version of (P5) for the MAC case by including the power allocations among the two users, $p_1$ and $p_2$, as optimization variables under the constraint $p_1+p_2\leq P$, and replacing $P_1,P_2$ with $p_1,p_2$. A suboptimal solution can be found for each problem via a similar AO algorithm as that for (P5) by iteratively optimizing $(r,p_1,p_2)$, $(r,\rho_\ff)$, or $(r,\phi_m^\cc)$ at each time with all the other variables being fixed. On the other hand, the capacity region outer bound $\mathcal{C}_{\BC,\OO}^\cc$ serves as an outer bound for $\mathcal{R}_{\BC,\ff}^\cc$ since $\mathcal{R}_{\BC,\ff}^\cc\subseteq \mathcal{C}_{\BC}^\cc$ \cite{Elements}, i.e., $\mathcal{R}_{\BC,\ff}^\cc\subseteq \mathcal{R}_{\BC,\ff,\OO}^\cc\overset{\Delta}{=}\mathcal{C}_{\BC,\OO}^\cc$.
		\begin{table*}[t]
			\centering
			\caption{Summary of Main Results for IRS-Aided Two-User BC}\label{table_BC}
			\resizebox{\textwidth}{!}{
				\begin{tabular}{|c|c|c|c|}
					\hline 
					&	Distributed	& Centralized &  Comparison (Twin Channels)\\ 
					\hline
					Capacity Region & $\mathcal{C}_{\BC}^\dd$ (duality) & \tabincell{c}{$\mathcal{C}_{\BC,\II}^\cc$ (rate-profile); $\mathcal{C}_{\BC,\OO}^\cc$ (duality)\\ $\mathcal{C}_{\BC,\II}^\cc\subseteq\mathcal{C}_{\BC}^\cc\subseteq \mathcal{C}_{\BC,\OO}^\cc$} & \tabincell{c}{
						$\mathcal{C}_{\BC}^\dd\subseteq \mathcal{C}_{\BC}^\cc$ (with $\bar{h}_k=0,\ \forall k$)} \\
					\hline
					\tabincell{c}{Achievable Rate\\ Region with TDMA} & $\mathcal{R}^\dd_{\BC,\ttt}$ (duality) & $\mathcal{R}^\cc_{\BC,\ttt}$ (duality) & $\mathcal{R}^\dd_{\BC,\ttt}\subseteq \mathcal{R}^\cc_{\BC,\ttt}$ \\
					\hline 
					\tabincell{c}{Achievable Rate\\ Region with FDMA} & \tabincell{c}{$\mathcal{R}^\dd_{\BC,\ff}$ (duality)\\ $\mathcal{R}^\dd_{\BC,\ttt}\subseteq\mathcal{R}^\dd_{\BC,\ff}\subseteq \mathcal{C}^\dd_{\BC}$ } & \tabincell{c}{$\mathcal{R}^\cc_{\BC,\ff,\II}$ (rate-profile); $\mathcal{R}^\cc_{\BC,\ff,\OO}$ \\$\mathcal{R}^\cc_{\BC,\ff,\II}\subseteq\mathcal{R}^\cc_{\BC,\ff}\subseteq \mathcal{R}_{\BC,\ff,\OO}^\cc$\\ $\mathcal{R}^\cc_{\BC,\ttt}\subseteq\mathcal{R}^\cc_{\BC,\ff}\subseteq \mathcal{C}^\cc_{\BC}$} & $\mathcal{R}_{\BC,\ff}^\dd\subseteq \mathcal{R}_{\BC,\ff}^\cc$ (with $\bar{h}_k=0,\ \forall k$)\\
					\hline
			\end{tabular}
		}
		
		\vspace{-3mm}
		\end{table*}
	
		\subsection{Performance Comparison for IRS-Aided Two-User BC}
		Finally, we extend the capacity/rate region comparison of the two deployment strategies in Section \ref{sec_comparison} for the MAC to its dual BC.
		\begin{proposition}\label{prop_compare_BC}
			Under $\bar{h}_1=\bar{h}_2=0$ and the twin channel condition in Assumption 1, the capacity region of the two-user BC under centralized IRS deployment contains that under the distributed IRS deployment, i.e., $\mathcal{C}^\dd_{\BC}\subseteq \mathcal{C}^\cc_{\BC}$.
		\end{proposition}
		\begin{IEEEproof}
			Recall from Proposition \ref{prop_compare} that $\mathcal{C}^\dd\subseteq \mathcal{C}^\cc$ holds for the two-user MAC under the channel assumptions. Hence, it follows from (\ref{C_dis_BC}) and (\ref{C_central_BC}) that $\mathcal{C}^\dd_{\BC}=\mathrm{Conv}\Big({\bigcup}_{(P_1,P_2):P_1+P_2= P}\mathcal{C}^\dd\Big)\subseteq \mathrm{Conv}\Big({\bigcup}_{(P_1,P_2):P_1+P_2= P}\mathcal{C}^\cc\Big)=\mathcal{C}^\cc_{\BC}$. This thus completes the proof of Proposition \ref{prop_compare_BC}.
		\end{IEEEproof}
		
		Similarly, under the twin channels, by leveraging the MAC-BC duality and the results in Section \ref{sec_comparison}, we have $\mathcal{R}_{\BC,\ttt}^\dd\subseteq \mathcal{R}_{\BC,\ttt}^\cc$ for any $\bar{h}_k$'s and $\mathcal{R}_{\BC,\ff}^\dd\subseteq \mathcal{R}_{\BC,\ff}^\cc$ for $\bar{h}_k=0,\forall k$. In Table \ref{table_BC}, we summarize our main results for the two-user BC.
		
		\vspace{-3mm}
		\section{Numerical Examples}\label{sec_num}
		In this section, we provide numerical examples to validate our analytical results. We set $M=30$, $M_1=M_2=15$ unless specified otherwise. Under a three-dimensional coordinate system, the AP is located at $(0,0,10)$ in meter (m), and the two users are located at $(\bar{d}_1,0,1)$ m and $(-\bar{d}_2,0,1)$ m, respectively, with $\bar{d}_k$ denoting the horizontal AP-user distance for user $k$. The IRS under centralized deployment is located at $(0,0,9)$ m, and the two IRSs under distributed deployment are located at $(\bar{d}_1,0,2)$ m and $(-\bar{d}_2,0,2)$ m, respectively. Thus, the IRS-AP distance under centralized deployment equals to the IRS-user distances under distributed deployment, and the IRS-user distances under centralized deployment equal to the corresponding IRS-AP distances under distributed deployment. The above setup is consistent to the twin channel condition in Assumption 1. We consider the independent and identically distributed (i.i.d.) Rayleigh fading channel model,\footnote{We assume that the elements in both distributed and centralized IRSs follow a uniform linear array (ULA) configuration with half-wavelength spacing, which usually leads to i.i.d. Rayleigh fading in an isotropic scattering environment \cite{Rayleigh}.} where the entries in $\{\bar{h}_k\}$, $\{\mv{h}_k^\cc\}$ and $\mv{g}^\cc$ are generated as independent CSCG random variables with zero mean and variance equal to the path loss of the corresponding link modeled as $\gamma=\gamma_0(1/d)^{\bar{\alpha}}$, where $\gamma_0=-30$ dB (corresponding to a carrier frequency of $755$ MHz), $d$ denotes the link distance in m, and $\bar{\alpha}$ denotes the path loss exponent. Note that our analytical results apply to arbitrary channels and due to the space limitation, we omit the numerical results for other channel models such as Rician fading. We set $\bar{\alpha}=3.5$ for the direct AP-user channels in $\{\bar{h}_k\}$ and $\bar{\alpha}=3$ for the reflected channels in $\{\mv{h}_k^\cc\}$ and $\mv{g}^\cc$. We also generate $\{\mv{h}_k^\dd\}$ and $\{\mv{g}_k^\dd\}$ similarly according to Assumption 1. The numbers of equally spaced points for approximating the rate ratio $\alpha_1$ and the power ratio $\alpha_\PP$ are set as $L=100$ and $L_\PP=100$, respectively.
		
		\begin{figure}[t]
			\centering
			\subfigure[Capacity region]{
				\includegraphics[width=8cm]{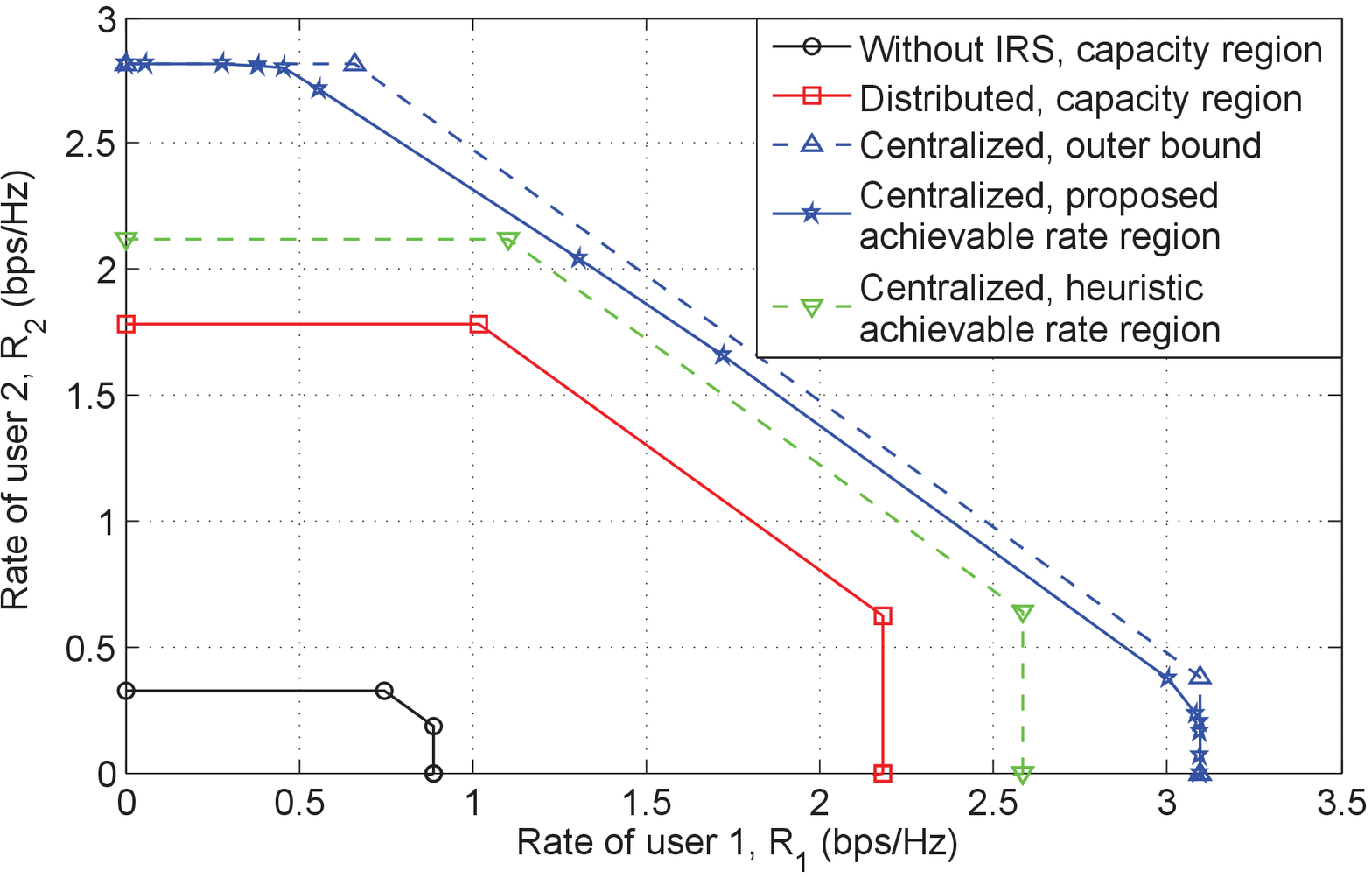}}
			\subfigure[Achievable rate regions with TDMA and FDMA]{
				\includegraphics[width=8cm]{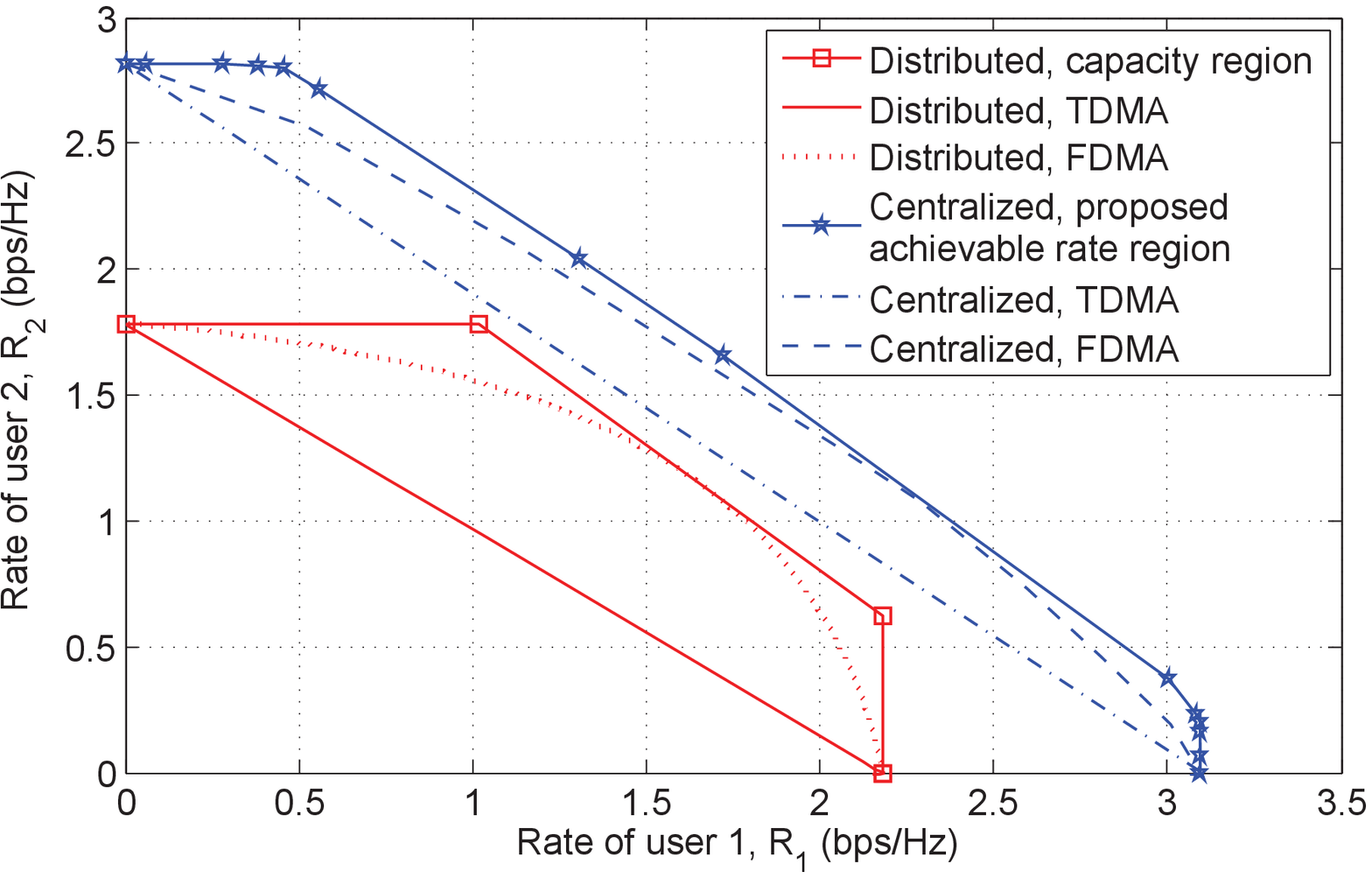}}
			\caption{Capacity/rate region comparison for IRS-aided two-user MAC under homogeneous user distance setup ($\bar{d}_1=\bar{d}_2=500$ m).}\label{fig_homo}
		\end{figure}
	
			\begin{figure}[t]
		\centering
		\subfigure[Capacity region]{
			\includegraphics[width=8cm]{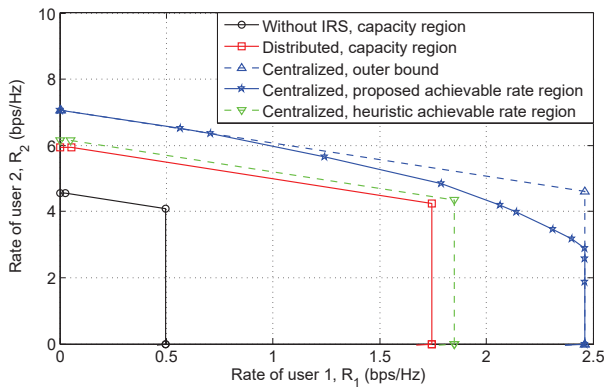}}
		\subfigure[Achievable rate regions with TDMA and FDMA]{
			\includegraphics[width=8cm]{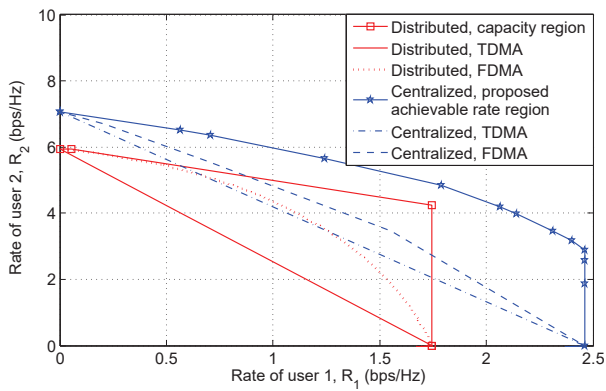}}
		\caption{Capacity/rate region comparison for IRS-aided two-user MAC under heterogeneous user distance setup ($\bar{d}_1\!\!=\!\!500$ m, $\bar{d}_2\!\!=\!\!200$ m).}\label{fig_hete}
	\end{figure}
	
		\subsection{IRS-Aided Two-User MAC}\label{sec_num_MAC}
		First, we focus on the two-user MAC investigated in Sections \ref{sec_sys}--\ref{sec_comparison}. We set $\frac{P_1}{\sigma^2}=\frac{P_2}{\sigma^2}=120$ dB and $Q=200$. First, we consider a \emph{homogeneous} user distance setup where both users have a horizontal distance of $\bar{d}_1=\bar{d}_2=500$ m from the AP, and randomly generate their channels based on i.i.d. Rayleigh fading. In Fig. \ref{fig_homo} (a), we show the capacity region for the traditional MAC without IRS and that with two distributed IRSs, as well as the outer and inner capacity region bounds with a centralized IRS. It is observed that the capacity region inner bound for centralized deployment contains the capacity region with distributed deployment, while the latter also contains the capacity region without IRS. This thus validates the effectiveness of deploying IRS in enlarging the capacity region as well as the advantage of centralized IRS deployment over distributed IRS deployment (even with the user-AP direct channels) under our assumed twin channel conditions. It is also interesting to observe that the capacity gain of centralized deployment over distributed deployment is more pronounced when the rates of the two users are \emph{asymmetric}, since the larger passive beamforming gain provided by the centralized IRS is more useful for the user with larger rate requirement. In addition, we show the achievable rate region under centralized deployment by a heuristic scheme with $\tilde{\mv{\Phi}}^\cc$ given in (\ref{phase1}) by setting $\theta_1=\theta_2=0$ (i.e., without the additional phase rotations designed for the two sub-surfaces to further align their reflected signals, as given in the proof of Proposition \ref{prop_compare}). The resultant achievable rate region is observed to be significantly smaller than our proposed one, which validates the efficacy of our proposed rate-profile based optimization. Furthermore, we show in Fig. \ref{fig_homo} (b) the achievable rate regions with TDMA and FDMA under the two deployment strategies. It is observed that for both deployment cases, FDMA outperforms TDMA, which is consistent with our analytical results; moreover, the achievable rate region of centralized deployment contains that of distributed deployment for both TDMA and FDMA cases, even when the direct channels are present. The above results indicate that centralized deployment outperforms distributed deployment in both the capacity-achieving non-orthogonal multiple access (NOMA) and practical OMA schemes.
				
		\begin{figure}[t]
	\subfigure[Maximum common rate vs. $\bar{d}_2$]{
		\includegraphics[width=8cm]{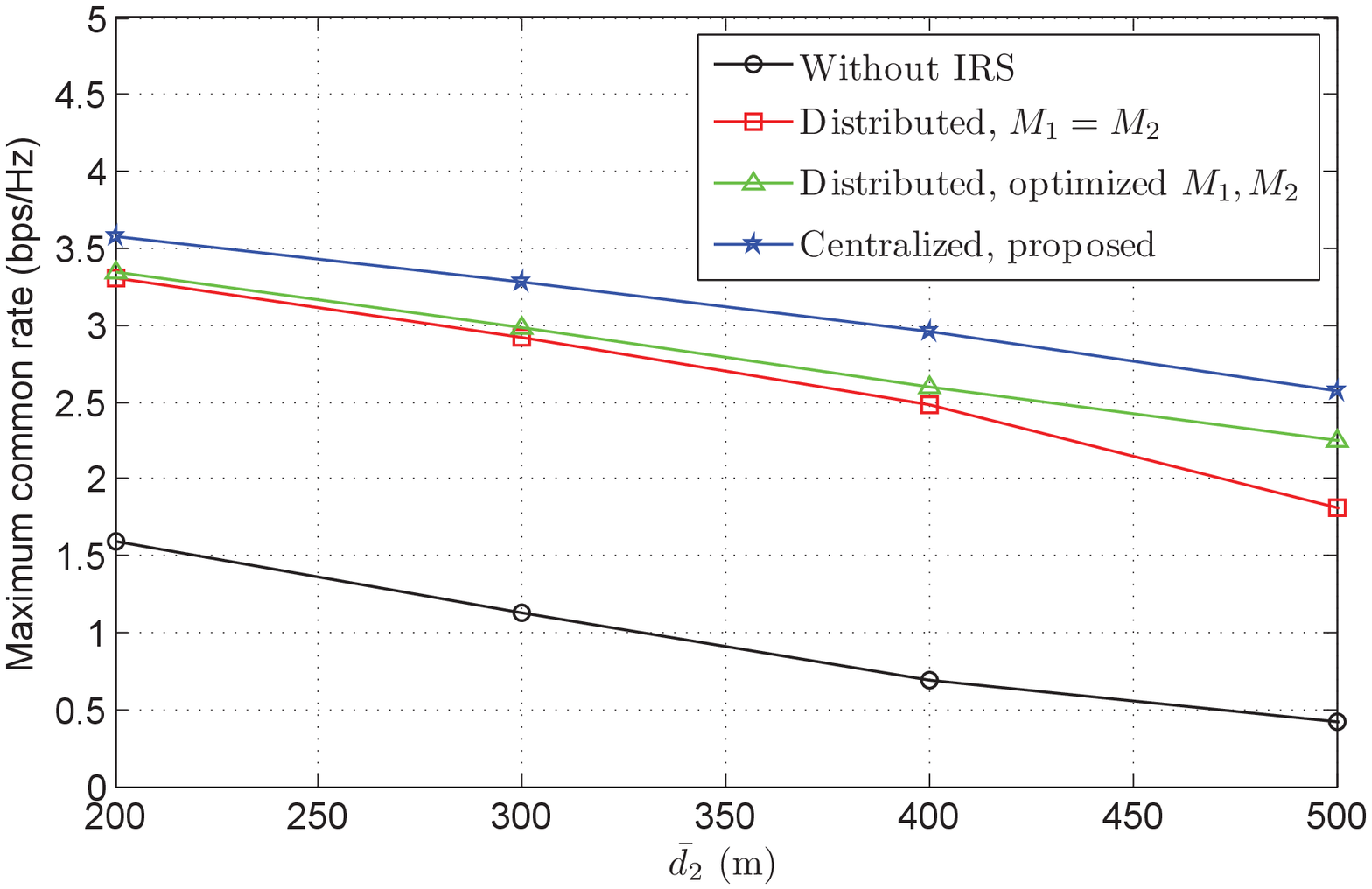}}
	\subfigure[\hbox{Maximum common rate vs. $M_2$ under distributed deployment}]{
		\includegraphics[width=8cm]{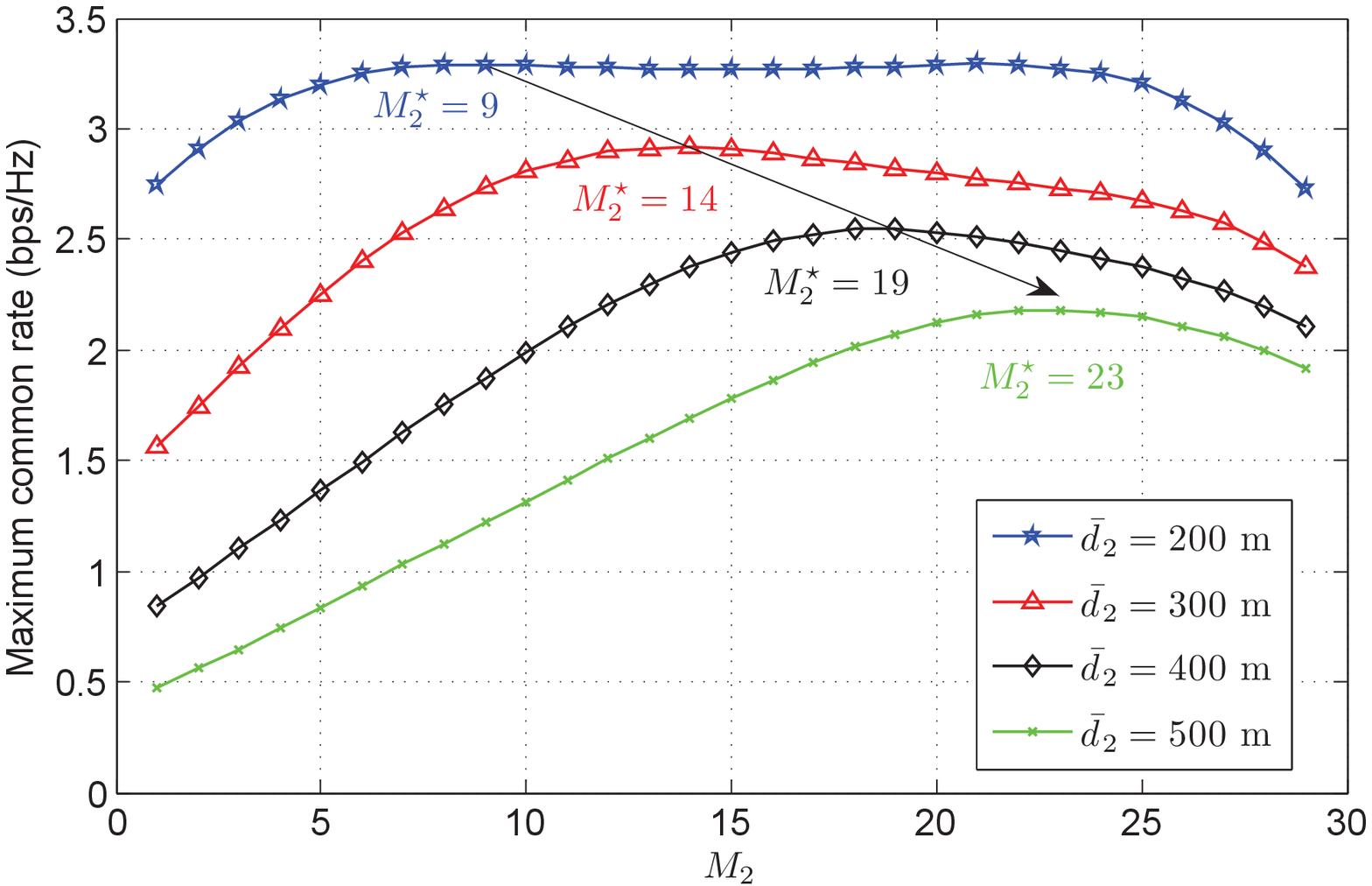}}
	\caption{Maximum common rate comparison for IRS-aided two-user MAC.}\label{fig_common}
\end{figure}
		
		Next, we consider a \emph{heterogeneous} user distance setup with $\bar{d}_1=500$ m and $\bar{d}_2=200$ m. In Fig. \ref{fig_hete} (a) and Fig. \ref{fig_hete} (b), we show the corresponding capacity region and achievable rate regions with TDMA/FDMA for different IRS deployment strategies, respectively. It is observed that the comparison results among different regions are similar as those under the homogeneous user distance setup shown in Fig. \ref{fig_homo}, as expected. However, it is observed that under the heterogeneous user distance setup, the performance gain of centralized deployment over distributed deployment is more pronounced for the user farther away from the AP, for both the NOMA and OMA schemes. This suggests that centralized deployment is more effective to alleviate the ``\emph{near-far}'' problem and yield more fair achievable rates for the users in the network. To show this benefit more clearly, we fix $\bar{d}_1=200$ m and show in Fig. \ref{fig_common} (a) the \emph{maximum common rate} achievable for the two users on the capacity regions of different deployment strategies versus (vs.) the horizontal AP-user distance for user $2$, $\bar{d}_2$, where the results are averaged over $100$ independent fading channel realizations. It is observed that as $\bar{d}_2$ increases from $200$ m (i.e., the ``near-far'' problem becomes more severe), the rate gain of the centralized deployment with our proposed design over distributed deployment becomes more prominent. Moreover, it is worth noting that the performance of distributed deployment can be further enhanced by optimizing the element allocations among the two IRSs, i.e., $M_1$ and $M_2$, which can be seen from Fig. \ref{fig_common} (b) on the maximum common rate versus $M_2$. It can be observed from Fig. \ref{fig_common} (b) that as $\bar{d}_2$ increases, the optimal number of elements for IRS 2, $M_2$, generally increases, i.e., more elements should be allocated to IRS 2 that serves the farther-away user 2. For example, when the two users have the same distance to the AP and thus similar path loss, the common rate is dominated by the interference from the second-decoded user on the first-decoded user, thus the two users should be allocated with slightly different numbers of elements to enhance the SIC performance (e.g., $M_2^\star=9$ or $21$ when $\bar{d}_2=200$ m); however, when user 2 moves away from the AP, it should be decoded secondly and the common rate is dominated by its own signal power, thus more elements should be placed in its vicinity (e.g., $M_2^\star=23$ when $\bar{d}_2=500$ m). We also show the maximum common rate with the optimized element allocations under the distributed deployment in Fig. \ref{fig_common} (a), which outperforms that with equal element allocations but is also outperformed by the centralized deployment, for which there is no issue of elements allocation.
			
				\begin{figure}[t]
			\centering
			\subfigure[Capacity region]{
				\includegraphics[width=8cm]{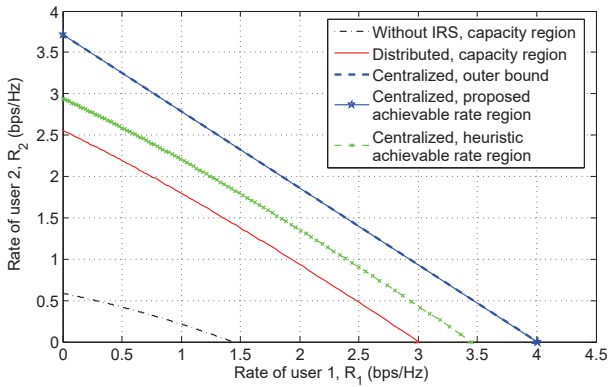}}
			\subfigure[Achievable rate regions with TDMA and FDMA]{
				\includegraphics[width=8cm]{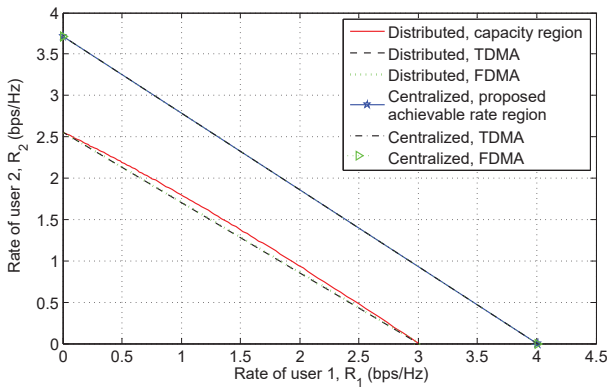}}
			\caption{Capacity/rate region comparison for IRS-aided two-user BC under homogeneous user distance setup ($\bar{d}_1=\bar{d}_2=500$ m).}\label{fig_BC_homo}
		\end{figure}
		
		\begin{figure}[t]
			\centering
			\subfigure[Capacity region]{
				\includegraphics[width=8cm]{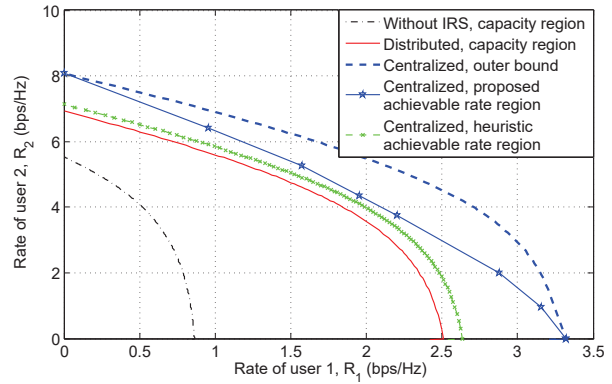}}
			\subfigure[Achievable rate regions with TDMA and FDMA]{
				\includegraphics[width=8cm]{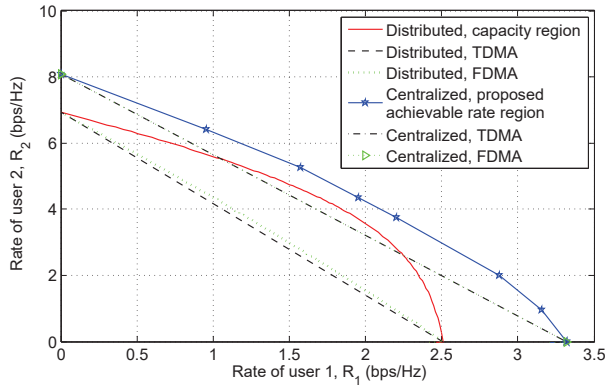}}
			\caption{Capacity/rate region comparison for IRS-aided two-user BC under heterogeneous user distance setup ($\bar{d}_1=500$ m, $\bar{d}_2=200$ m).}\label{fig_BC_hete}
		\end{figure}
		\vspace{-3mm}
		\subsection{IRS-Aided Two-User BC}
		Finally, we consider the two-user BC under the setup of $\frac{P}{\sigma^2}=123$ dB and $Q_{\BC}=200$, where the channels are dual to the two-user MAC considered above. In Fig. \ref{fig_BC_homo} and Fig. \ref{fig_BC_hete}, we show the capacity and rate regions of various schemes under a homogeneous user distance setup with $\bar{d}_1=\bar{d}_2=500$ m and a heterogeneous user distance setup with $\bar{d}_1=500$ m, $\bar{d}_2=200$ m, respectively. It is observed that for both setups, our proposed achievable rate region under centralized deployment contains the capacity region under distributed deployment, while the latter also contains the capacity region without the IRS; moreover, the relationships between different capacity/rate regions are also consistent with our analytical results in Section \ref{sec_BC}. For comparison, we show a heuristic achievable rate region for the centralized deployment case by taking the convex hull of the union set of the heuristic achievable rate regions for its dual MAC shown in Section \ref{sec_num_MAC} with different user transmit powers, which is observed to be substantially smaller than our proposed one based on rate-profile. In addition, the rate gain of centralized deployment is more pronounced for the farther away user from the AP under the heterogeneous distance setup, which is consistent with our results for the MAC case as discussed above. Furthermore, by comparing Fig. \ref{fig_BC_homo} (b) and Fig. \ref{fig_BC_hete} (b), it is observed that for both deployment strategies, the achievable rate regions with TDMA and FDMA approach the capacity region quite well under the homogeneous distance setup, while they are strictly suboptimal under the heterogeneous distance setup. This is because when the average channel gains of the two users are similar, the overall BC capacity region with optimized IRS reflection coefficients is approximately a triangle, which can be achieved by both TDMA and FDMA with the same IRS reflection coefficients through time sharing among the two users.
		
						\begin{table*}[t]
			\centering
			\caption{Summary of $\Delta\lambda$}\label{table_prop_relax}
			\resizebox{\textwidth}{!}{\begin{tabular}{|c|c|c|c|}
					\hline 
					\multicolumn{3}{|c|}{ }  &	$\Delta\lambda$	\\ 
					\hline
					\multicolumn{3}{|c|}{$\mathfrak{Re}\{f_{2,\pi_1m}\phi_m^\cc \}\geq 0$, $\mathfrak{Re}\{f_{2,\pi_2m}\phi_m^\cc \}\geq 0$} & $0$ \\
					\hline
					\multicolumn{3}{|c|}{$\mathfrak{Re}\{f_{2,\pi_1m}\phi_m^\cc \}< 0$, $\mathfrak{Re}\{f_{2,\pi_2m}\phi_m^\cc \}< 0$} & $\pi$ \\
					\hline
					\multirow{6}*{ \tabincell{c}{$\mathfrak{Re}\{f_{2,\pi_im}\phi_m^\cc \}\geq 0$, $\mathfrak{Re}\{f_{2,\pi_jm}\phi_m^\cc \}\leq 0$,\\ $i,j\in \{1,2\}$, $i\neq j$}}  &  \multicolumn{2}{|c|}{$\eta_i+\lambda\in [0,\frac{\pi}{2}]$, $\eta_j+\lambda \in (\frac{\pi}{2},\pi]$}  & $-\arccos\{b\}$ \\
					\cline{2-4} & \multirow{2}*{$\eta_i+\lambda\in [0,\frac{\pi}{2}]$, $\eta_j+\lambda \in [\pi,\frac{3\pi}{2})$} & $\eta_j-\pi<\eta_i$ & $-\eta_i-\lambda-\arccos\{b\cos(\eta_i+\lambda)\}$ \\
					\cline{3-4} & & $\eta_j-\pi\geq\eta_i$ & $-\eta_i-\lambda+\arccos\{b\cos(\eta_i+\lambda)\}$
					\\ \cline{2-4}
					&  \multicolumn{2}{|c|}{$\eta_i+\lambda\in [\frac{3\pi}{2},2\pi)$, $\eta_j+\lambda \in [\pi,\frac{3\pi}{2})$}  & $\arccos\{b\}$ \\
					\cline{2-4} & \multirow{2}*{$\eta_i+\lambda\in [\frac{3\pi}{2},2\pi)$, $\eta_j+\lambda \in (\frac{\pi}{2},\pi]$} & $\eta_j+\pi\geq\eta_i$ & $-\eta_i-\lambda+\arccos\{b\cos(\eta_i+\lambda)\}$ \\
					\cline{3-4} & & $\eta_j+\pi<\eta_i$ & $-\eta_i-\lambda-\arccos\{b\cos(\eta_i+\lambda)\}$
					\\
					\hline
			\end{tabular}}
			
			\vspace{-3mm}
		\end{table*}

\section{Concluding Remarks}\label{sec_con}
This paper studied the capacity region of an IRS-aided two-user communication system, under two practical IRS deployment strategies. For the uplink IRS-aided MAC, we characterized the capacity region and achievable rate regions with TDMA/FDMA for both deployment strategies, and proved that the regions under centralized IRS deployment contain the corresponding ones under distributed IRS deployment, assuming a practical ``twin channel'' setup. The results were also extended to the downlink IRS-aided BC by leveraging the MAC-BC duality, where the performance gain of centralized over distributed IRS deployment was proved to be also valid. Numerical results validated our analysis and revealed that the superiority of centralized over distributed IRS deployment is more prominent when the two users have asymmetric rate requirements and/or channel conditions. 
		
It is worth noting that the established rate-profile based framework for capacity region characterization can be readily extended to the more general IRS-aided $K$-user MAC and BC with arbitrary $K$, by considering all the possible decoding orders among the users. Besides, there are other appealing directions worth pursuing in future works. First, besides MAC and BC studied in this paper, it is interesting to extend our proposed framework for characterizing the capacity region of IRS-aided channels to other channel models such as the interference channel (IC), multicast channel, etc. Moreover, it is worthwhile to compare the performance of distributed versus centralized IRS deployment under more general network models such as the multi-cell multi-user network \cite{Multicell}, where the complicated interplay between the inter-cell and intra-cell interference needs to be judiciously considered in the IRS deployment design. Finally, it is worth mentioning that the IRS deployment problem is addressed in this paper from an information-theoretic viewpoint, while from an implementation perspective, other practical factors may also need to be considered, such as the backhaul cost for information exchange (e.g., CSI), site/space constraint, availability of line-of-sight (LoS) channels, etc. Furthermore, the results in this paper are based on the assumption of identical channel distribution for both distributed and centralized IRS cases, while in practice, the channel distribution may vary at different IRS locations due to distinct terrain features and as a result, the performance comparison between the two IRS deployment strategies is more involved.
		
\vspace{-3mm}
\appendix
\subsection{Proof of Proposition \ref{prop_centralized}}\label{proof_prop_centralized}
Let $\mathcal{C}^{\cc^\star}$ denote the right-hand side (RHS) of (\ref{CR_cen}). First, the achievability of $\mathcal{C}^{\cc^\star}$ (i.e., $\mathcal{C}^{\cc^\star}\subseteq \mathcal{C}^{\cc}$) is evident from the problem formulation of (P1). Next, we prove the converse, i.e., $\mathcal{C}^{\cc^\star}$ is an outer bound of ${\mathcal{C}}^\cc$. Specifically, for any given $\mv{\alpha}$ and $\mv{\Phi}^\cc$, let $r_{\II}^\star(\mv{\alpha},\mv{\Phi}^\cc)$ and $r_{\II\II}^\star(\mv{\alpha},\mv{\Phi}^\cc)$ denote the optimal value of (P1) with $\mv{\pi}=\mv{\pi}^{\II}$ and $\mv{\pi}=\mv{\pi}^{\II\II}$, respectively, and $(R_1^{\cc^\star}(\mv{\alpha},\mv{\Phi}^\cc),R_2^{\cc^\star}(\mv{\alpha},\mv{\Phi}^\cc))$ denote the Pareto-optimal rate-pair along $\mv{\alpha}$ defined similarly as $(R_1^{\cc^\star}(\mv{\alpha}),R_2^{\cc^\star}(\mv{\alpha}))$. We then have $\mathcal{C}^\cc(\mv{\Phi}^\cc)=\mathrm{Conv}\big((0,0){\bigcup}_{\mv{\alpha}:\alpha_1\in [0,1]} (R_1^{\cc^\star}(\mv{\alpha},\mv{\Phi}^\cc),R_2^{\cc^\star}(\mv{\alpha},\mv{\Phi}^\cc))\big)$
and 
\begin{align}
		&{\mathcal{C}}^\cc=\mathrm{Conv}\Big({\bigcup}_{{\mv{\Phi}}^\cc\in \mathcal{F}^\cc} {\mathcal{C}}^\cc({\mv{\Phi}}^\cc)\Big)\nonumber\\
		=&\mathrm{Conv}\big((0,0)\ \underset{{\mv{\Phi}}^\cc\in \mathcal{F}^\cc}{\bigcup}\underset{\mv{\alpha}:\alpha_1\in [0,1]}{\bigcup} (R_1^{\cc^\star}(\mv{\alpha},\mv{\Phi}^\cc),R_2^{\cc^\star}(\mv{\alpha},\mv{\Phi}^\cc))\big)\nonumber\\
		\subseteq&{\mathcal{C}}^{\cc^\star},
\end{align}
since $R_1^{\cc^\star}(\mv{\alpha})\geq R_1^{\cc^\star}(\mv{\alpha},\mv{\Phi}^\cc)$ and $R_2^{\cc^\star}(\mv{\alpha})\geq R_2^{\cc^\star}(\mv{\alpha},\mv{\Phi}^\cc)$ hold for any $\mv{\Phi}^\cc$. This completes the proof of the converse part. Consequently, we have $\mathcal{C}^{\cc}= \mathcal{C}^{\cc^\star}$ and Proposition \ref{prop_centralized} is thus proved.
		
\vspace{-3mm}
\subsection{Proof of Proposition \ref{prop_relax}}\label{proof_prop_relax}
We prove Proposition \ref{prop_relax} by showing that for any feasible solution ${\phi}_m^\cc$ to (P3-m) with $|\phi_m^\cc|<1$, we can always construct a new solution $\bar{\phi}_m^\cc$ with $|\bar{\phi}_m^\cc|=1$ that yields a no smaller objective value of (P3-m). Specifically, define $f_{2,\pi_1m}=a_1e^{j\eta_1}$ and $f_{2,\pi_2m}=a_2e^{j\eta_2}$ with $a_1,a_2\geq 0$, $\eta_1,\eta_2\in [0,2\pi)$; $\phi_m^{{\rm C}}=be^{j\lambda}$ with $0\leq b<1$, $\lambda\in [0,2\pi)$; and $\bar{\phi}_m^{{\rm C}}=e^{j(\lambda+\Delta \lambda)}$ with $|\bar{\phi}_m^{{\rm C}}|=1$, $\Delta\lambda\in [0,2\pi)$. We show that we can always find a $\Delta \lambda$ such that  $\mathfrak{Re}\{f_{2,\pi_1m}\bar{\phi}_m^{{\rm C}}\}\geq \mathfrak{Re}\{f_{2,\pi_1m}\phi_m^{{\rm C}}\}$ and $\mathfrak{Re}\{f_{2,\pi_2m}\bar{\phi}_m^{{\rm C}}\}\geq \mathfrak{Re}\{f_{2,\pi_2m}\phi_m^{{\rm C}}\}$ hold, and consequently, the objective value of (P3-m) with $\bar{\phi}_m^\cc$ is no smaller than that of ${\phi}_m^\cc$ since $\beta^{\frac{1}{1-\alpha_{\pi_1}}}-\beta$ and $\beta-1$ are non-decreasing and increasing functions of $\beta$, respectively. Due to the space limitation, we summarize the choice of such $\Delta\lambda$ for different cases in Table \ref{table_prop_relax}, for which the detailed derivations are omitted for brevity.
		
\vspace{-3mm}
\subsection{Proposed Solution to (P5)}\label{prop_P5}
Note that with given $\mv{\Phi}^\cc$, (P5) can be shown to be a convex optimization problem over $(r,\rho_\ff)$. On the other hand, with given $\rho_\ff$ and $\{\phi_i,i\neq m\}_{i=1}^M$, (P5) is equivalent to
\begin{align}
		\mbox{(P5-m)} &\nonumber\\
				\! \underset{r,\phi_m^\cc: |\phi_m^\cc|=1}{\mathtt{max}}\!\!\!\! &r\\
			\mathtt{s.t.}\quad &2\mathfrak{Re}\{f_{2,1m}\phi_m^\cc\}+f_{1,1m}\geq \frac{(2^{\frac{\alpha_1 r}{\rho_\ff}}-1)\rho_\ff\sigma^2}{P_1}\label{P5mc1}\\
			&2\mathfrak{Re}\{f_{2,2m}\phi_m^\cc\}+f_{1,2m}\geq \frac{(2^{\frac{(1\!-\!\alpha_1) r}{1\!-\!\rho_\ff}}-1)(1\!-\!\rho_\ff)\sigma^2}{P_2}.\label{P5mc2}
\end{align} 
where $f_{1,km}$ and $f_{2,km}$ are defined in Section \ref{sec_cen_inner}. Similar to (P3-m), (P5-m) can be shown to be equivalent to its relaxed version with $|\phi_m^\cc|=1$ replaced by $|\phi_m^\cc|\leq 1$, which is a convex optimization problem and can be solved efficiently via the interior-point method. Hence, by iteratively optimizing $(r,\rho_\ff)$ or $(r,\phi_m^\cc)$ for one $m\in \mathcal{M}$ with all the other variables being fixed at each time, we can obtain a feasible solution to (P5) denoted by $(\tilde{r}(\mv{\alpha}),\tilde{\rho}_\ff(\mv{\alpha}),\tilde{\mv{\Phi}}^\cc(\mv{\alpha}))$. Based on this, an inner bound of $\mathcal{R}_\ff^\cc$ can be obtained \hbox{as $\mathcal{R}_{\ff,\II}^\cc=\mathrm{Conv}\Big((0,0) {\bigcup}_{\mv{\alpha}:\alpha_1\in [0,1]}(\alpha_1,1-\alpha_1)\tilde{r}(\mv{\alpha})\Big)\subseteq \mathcal{R}_\ff^\cc$.}
		
\vspace{-3mm}
\subsection{Proof of Lemma \ref{lemma_phase}}\label{proof_lemma_phase}
For ease of exposition, we assume that $c_2\geq c_1$ without loss of generality. For the case of $|c_2-c_1|\geq \pi$ and $\theta=\pi/2-\min(c_1,c_2)$, we have $\cos(c_1+\theta)=\cos(\pi/2)=0$, and $c_2+\theta=\pi/2+c_2-c_1\in [3\pi/2,5\pi/2]$, thus  $\cos(c_2+\theta)\geq0$. On the other hand, for the case of $|c_2-c_1|< \pi$ and $\theta=\pi/2-\max(c_1,c_2)$, we have $\cos(c_2+\theta)=\cos(\pi/2)=0$, and $c_1+\theta=\pi/2+c_1-c_2\in (-\pi/2,\pi/2]$, thus $\cos(c_1+\theta)\geq 0$. Therefore, we have  $\cos(c_1+\theta)\geq 0$ and $\cos(c_2+\theta)\geq 0$ for both cases, and thus $|a_k+b_ke^{j\theta}|=(|a_k|^2+|b_k|^2+2|a_k||b_k|\cos(c_k+\theta))^{\frac{1}{2}}\geq |a_k|$ holds for both $k=1$ and $k=2$. This thus completes the proof of Lemma \ref{lemma_phase}.
		
\bibliographystyle{IEEEtran}
\bibliography{IRS_Deployment_Final}		
\end{document}